\documentclass[
superscriptaddress,
nofootinbib,
nobibnotes,
 amsmath,amssymb,
prb,onecolumn
]{revtex4-1}
\usepackage{graphicx}
\usepackage{dcolumn}
\usepackage{bm}
\usepackage{notoccite}
\usepackage{physics}
\usepackage{xcolor}
\usepackage{mathtools}
\begin{document}

\title{Nonlinear Hall effect with time-reversal symmetry: \\ Theory and material realizations}

\author{Carmine Ortix}
\affiliation{Institute for Theoretical Physics, Center for Extreme Matter and Emergent Phenomena, Utrecht University, Princetonplein 5, 3584 CC Utrecht,  Netherlands}
\affiliation{Dipartimento di Fisica ``E. R. Caianiello", Universit\'a di Salerno, IT-84084 Fisciano, Italy}

\begin{abstract}
The appearance of a Hall conductance necessarily requires breaking of time-reversal symmetry, either by an external magnetic field or by the internal magnetization of a material. However, as a second response, Hall dissipationless transverse currents can appear even in time-reversal symmetric conditions  in non-centrosymmetric materials. Moreover, this non-linear effect has a quantum origin: it is related to the geometric properties of the electronic wavefunctions and encoded in the dipole moment of the Berry curvature.  Here we review the general theory underpinning this effect and discuss various material platforms where non-linear Hall transverse responses have been theoretically proposed and experimentally verified. 
On the theoretical front, the link between the non-linear Hall effect and the Berry curvature dipole is discussed using Boltzmann transport theory. 
On the material front, different platforms, including topological crystalline insulators, transition metal dichalcogenides, graphene, and Weyl semimetals are reviewed. 
\end{abstract}

\maketitle

\section{Introduction}
The electric Hall effect -- the phenomenon originally observed by Edwin H. Hall in 1879 -- is the production of a transverse voltage in current carrying conductors subject to an external out-of-plane magnetic field. 
Studies of the Hall effect have led to important discoveries in condensed matter physics. These include the quantization of the Hall conductance~\cite{kli80} and its interpretation in terms of the topological properties of the Landau levels~\cite{tho82}, which 
have subsequently led to the discovery of different topological phases of matter~\cite{has10,qi11}.
The relation between topology and the Hall effect is due to the fact that the quantum wavefunctions of the electrons have a geometric structure that yields an effective bending 
in time-reversal symmetry-broken materials. In ferromagnetic conductors this bending leads to the intrinsic anomalous Hall effect (AHE). Although the AHE has been originally observed by Hall himself and the first underpinning microscopic theories developed in the 50's by Karplus and Luttinger~\cite{kar54,lut58}, the fundamental understanding of its intrinsic part in terms of concepts based on geometry and topology has been achieved only recently~\cite{jun02,Haldane2}. Since then, 
an important area of research has concerned the identification of magnetic quantum materials displaying a sizeable AHE both in its unquantized~\cite{nag10} and quantized version~\cite{Haldane,cha13}. 

Hall effects cannot instead probe the non-trivial geometric properties of the Bloch states in time-reversal invariant materials with broken inversion symmetry. This is because the Berry curvature of a system with time-reversal symmetry is an odd function of the crystal momentum ${\bf k}$. Its integral weighed by the equilibrium Fermi distribution function is then forced to vanish: Kramers' theorem ensures that states at ${\bf k}$ and $- {\bf k}$ are either both populated or not populated. It has been recently established~\cite{dey09,moo10,Sodemann2015}, however, that this assertion holds true only as long as the linear response regime is considered, {\it i.e.} for Hall voltages that are linearly proportional to the external electric field. Non-linear Hall currents that are quadratic in the external electric field can appear even in time-reversal symmetric conditions. This non-linear effect has an intrinsic contribution directly related to a quantity that can be seen as the dipole moment of the Berry curvature~\cite{Sodemann2015}. The non-linear Hall effect (NLHE) due to the Berry curvature dipole (BCD) has a quantum origin precisely as the intrinsic AHE, but cannot be quantized since it always requires a finite Fermi surface. 
It is also related to photogalvanic effects -- where the geometric properties of the wavefunctions also come into play-- involving  interband transitions at optical frequencies~\cite{dej17,mor16}, but with the important difference that the NLHE is governed by intraband processes involving electrons near the Fermi surface. 

This characteristic makes the NLHE due to BCD not only a unique toolbox to probe the geometric properties of electronic wavefunctions in time-reversal invariant materials. 
New quantum technologies could directly make use of this effect. The quadratic transverse current-voltage characteristic could be in fact used to rectify~\cite{iso20} and detect terahertz radiation~\cite{zha21} directly. In addition, such Hall rectifiers would have a number of advantages including the simplicity of the device architecture that would facilitate integration on-chip and mass production. 

The aim of this work is to review the fundamental theory and the material realizations of this recently discovered quantum effect. We have structured this review as follows. In Sec.~\ref{sec:theory} we discuss the theory underpinning the NLHE with time-reversal symmetric conditions using Boltzmann transport theory. In Sec.~\ref{sec:2dsoc}  we discuss the appearance of this effect in two-dimensional quantum materials characterized by a strong spin-orbit coupling. In Sec.~\ref{sec:nosoc2d} we review the mechanism responsible for the NLHE in spin-orbit free two-dimensional graphene. In Sec.~\ref{sec:3dmaterials} we consider the realization of this effect in bulk non-centrosymmetric three-dimensional crystals. Finally, in Sec.~\ref{sec:conc} we draw our conclusions.

 \section{Theory of the non-linear Hall effect in time-reversal symmetric conditions} 
 \label{sec:theory}

 \subsection{Symmetry constraints on the non-linear Hall effect}
  \label{subsec:thsymm}
   In this section we present, following Ref.~\cite{nan19}, a general symmetry analysis that  highlights the primary role played by crystalline symmetries in the occurrence of the NLHE with time-reversal symmetry. 
 As a warmup, let us first consider the constraints crystalline symmetries pose on the conventional Hall effect found in the linear response regime. Generally speaking, the electric current response to an external electric field is captured by the linear conductivity tensor defined by 
  \begin{equation*}
  j_{\alpha}= \sigma_{\alpha \, \beta} E_{\beta}.
  \end{equation*}
  The off-diagonal tensor components regulating transverse currents can be separated into symmetric parts that appear in systems with sufficiently low symmetry, usually when $\alpha,\beta$ do not correspond to principal crystallographic directions~\cite{wu17}, and a Hall antisymmetric part that explicitly requires time-reversal symmetry to be broken as a consequence of Onsager reciprocity relations~\cite{ons31}. In two-dimensional systems, the single antisymmetric conductivity tensor component that corresponds to the Hall conductivity can be written as 
  \begin{equation}
  \sigma_H = \dfrac{\epsilon_{\alpha \beta} \sigma_{\alpha \beta}}{2}, 
  \label{eq:sigmaH}
  \end{equation}
  with $\epsilon_{\alpha \beta}$ the two-dimensional Levi-Civita symbol. 
  The constraints on $\sigma_H$ due to crystalline symmetries can be identified by first recalling that  a generic point group symmetry is represented by an orthogonal matrix $\mathcal{O}$. Since the current and the electric fields transform as vectors under a generic coordinate change, the conductivity tensor must transform according to ${\mathcal O}^T \sigma \, {\mathcal O}$. This transformation rule implies that the antisymmetric Hall conductivity transforms as a pseudoscalar 
  \begin{equation*}
  \sigma_H= \det{\mathcal O} \sigma_H. 
  \end{equation*}
 As a consequence of the equation above, already the presence of a single point-group symmetry with $\det{\mathcal O}=-1$, such as a mirror line, forces the Hall conductance to vanish~\cite{fan12}. 
 
 Let us now consider the current response at second-order in electric fields. This defines a non-linear conductivity tensor
 \begin{equation*}
 j_{\alpha} = \chi_{\alpha \beta \gamma} E_{\beta} E_{\gamma}.
 \end{equation*} 
In strict analogy with the linear response regime, we can separate the non-linear tensor $\chi_{\alpha \beta \gamma}$ into components that contribute to the electrical power dissipation $j_{\alpha} E_{\alpha}$,  and dissipationless Hall components, which, contrary to $\sigma_H$, are not forced to vanish only by time-reversal invariance.  
This Hall conductance can be identified by 
antisymmetrizing the first index with either the second or the third -- both choices are equivalent because the tensor is by construction symmetric in the last two indices. 
In the particular case of two-dimensional systems, there are thus two independent components of the non-linear dissipationless Hall conductance reading 
\begin{equation}
\chi_{H \, \gamma} = \dfrac{\epsilon_{\alpha \beta} \chi_{\alpha \beta \gamma} }{2}, 
\label{eq:chiH}
\end{equation}
and transform as a pseudovector under a point group symmetry, {\it i.e.} 
\begin{equation*} 
\chi_{H \, \beta} = \det{\mathcal{O}} {\mathcal O}_{\beta \alpha} \chi_{H \, \alpha}. 
\end{equation*} 
The presence of a mirror line in this case only forces the non-linear Hall pseudovector to be orthogonal to the mirror plane. Therefore, $\chi_{H}$ can be non-vanishing. On the contrary, the presence of two or more mirror lines require 
$\chi_H$ to correspond to the null vector. 
Note also that the different transformation properties of the linear (Eq.~\ref{eq:sigmaH}) and non-linear (Eq.~\ref{eq:chiH}) conductivity tensors imply that in certain crystalline structures non-linear dissipationless Hall currents can exist in the complete absence of linear Hall currents, even if time-reversal is explicitly broken. As an example, in crystals with ${\mathcal C}_v$ point group symmetry the linear Hall conductance $\sigma_H$ vanishes independent of the presence of time-reversal symmetry. On the contrary $\chi_H$ will be only orthogonal to the unique mirror line of the crystal. 

 Following similar arguments, it is possible to derive the symmetry constraints on the (non)linear conductivity tensor in a three-dimensional material. The three independent components of the linear Hall conductance given by $\sigma_{H \, \gamma}=\epsilon_{\gamma \alpha \beta} \sigma_{\alpha \beta} /2$  transform as a pseudovector. Therefore, precisely as the non-linear Hall conductance of two-dimensional systems, the Hall vector of a three-dimensional material must be normal to any mirror plane. The presence of two independent mirror planes therefore forces all components of $\sigma_{H \, \gamma}$ to vanish. The three-dimensional non-linear Hall conductivity has instead nine independent components that transform as a rank-two pseudotensor. 
This tensor can be decomposed into a symmetric and an antisymmetric part~\cite{Sodemann2015}. The antisymmetric part transforms as a vector, and is potentially non-vanishing in the ten polar point groups ${\mathcal C}_n$ and ${\mathcal C}_{n v}$ where $n=1,2,3,4,6$. In this case, the polar axis determines the direction of the non-linear Hall vector. The existence of a finite symmetric part of the three-dimensional non-linear Hall conductance is instead strongly dependent on the presence of mirror symmetries. All non-centrosymmetric crystal point groups without left-handed symmetries can potentially have a non-vanishing nonlinear Hall conductance. The only crystal point groups with mirror symmetries that allow for a non-vanishing symmetric part of the non-linear Hall tensor are instead ${\mathcal C}_{1v}$, ${\mathcal C}_{2v}$ and ${\mathcal S}_4$. 
As we will see in the next section, the non-vanishing non-linear Hall conductance can be linked to the BCD using a Boltzmann equation approach in the constant relaxation time approximation. 
  
 \subsection{Non-linear Hall effect due to Berry curvature dipole} 
 \label{subsec:thbcd}
  We now employ the semiclassical Boltzmann transport framework in a simple single band model to relate the non-linear Hall conductance introduced in the preceding section to the BCD. We start by recalling that in the absence of externally applied magnetic fields, the semiclassical equations of motion accounting for the anomalous velocity term~\cite{sun99,xia10,nag10} can be written as 
\begin{eqnarray} 
\dot{\bf r}&=& \dfrac{1}{\hbar}  \nabla_{\bf k} \epsilon({\bf k}) + \dfrac{e}{\hbar}~{\bf E} \times {\bf \Omega}_{\bf k} \nonumber \\
\dot{\bf k}&=&\dfrac{1}{\hbar} e~{\bf E} 
\label{eq:semiclassicalmotion}
\end{eqnarray}
where $\epsilon$ is the energy dispersion of the metal in question, ${\bf E}$ is the driving electric field, and ${\bf \Omega}_{\bf k}$ is the Berry curvature defined as $\Omega_{k_a}= \epsilon_{abc} \partial_{k_b} {\mathcal A}_{k_c}$ with ${\mathcal A}_{k_c}\equiv -i \bra{u_{\bf k}}\partial_{k_c}\ket{u_{\bf k}} $ the Berry connection associated to the Bloch waves $\ket{u_{\bf k}}$ with crystalline momentum ${\bf k}$. To proceed further, we use that the electronic distribution function $f_{{\bf k}, {\bf r}, t}$ satisfies the semiclassical Boltzmann equation 
\begin{equation}
\partial_t  f_{{\bf r}, {\bf k}, t} + \dot{\bf k} \cdot {\bf \nabla}_{\bf k} f_{{\bf r}, {\bf k}, t}  +  \dot{\bf r} \cdot {\bf \nabla}_{\bf r} f_{{\bf r}, {\bf k}, t} = {\mathcal I}_{coll}\left\{f \right\}
\label{eq:boltzmanngen}
\end{equation} 
where the collision integral in the relaxation time approximation reads 
\begin{equation*}
{\mathcal I}_{coll}\left\{f \right\}= - \dfrac{\left(f_{{\bf r}, {\bf k}, t} - f^0_{{\bf k}}\right)}{\tau}.
\end{equation*} 
In the equation above $\tau$ indicates an intraband scattering time and $f^0_{{\bf k}}$ the equilibrium distribution function. We are interested in a stationary and homogeneous solution to the Boltzmann equation. Using Eqs.~\ref{eq:semiclassicalmotion},~\ref{eq:boltzmanngen} we have  
\begin{equation*}
\dfrac{e}{\hbar} \tau {\bf E} \cdot {\bf \nabla}_{\bf k} f_{{\bf k}} = - \left(f_{{\bf k}} - f^0_{{\bf k}}\right). 
\end{equation*}
To proceed further, we expand the distribution function in a series as $f_{{\bf k}}=f^0_{{\bf k}} + f^1_{{\bf k}} +  f^2_{{\bf k}} + \ldots$ where the term $f^n_{{\bf k}}$ is understood to vanish to the ${\bf E}^n$ order. 
One then finds the following structure for the linear and non-linear terms in the distribution function 
\begin{eqnarray} 
f_{{\bf k}}^1&=& - \dfrac{e}{\hbar}~\tau~E_{\alpha} \partial_{k_{\alpha}} f_{{\bf k}}^0 \nonumber \\ 
f_{{\bf k}}^2&=& \dfrac{e^2}{\hbar^2}~\tau^2E_{\alpha}E_{\beta}~\partial_{k_{\alpha}}\partial_{k_{\beta}} f_{{\bf k}}^0. 
\label{eq:intrinsicboltz}
\end{eqnarray}
We can now compute the current using the relation ${\bf j} = e \int d^d {\bf k} / (2 \pi)^d \dot{\bf r} f_{{\bf k}}$. 
At linear order in the electric field we have 
\begin{equation} 
j_{\alpha} = \dfrac{e^2 \tau}{\hbar^2} \int \dfrac{d^d k}{(2 \pi)^d} \partial_{k_{\alpha}}  \epsilon({\bf k}) \partial_{k_{\beta}}\epsilon({\bf k})~\left(- \dfrac{\partial f_{{\bf k}}^0}{\partial \epsilon}\right)~E_{\beta} + \dfrac{e^2}{\hbar}  \int \dfrac{d^d k}{(2 \pi)^d}  \epsilon_{\alpha \beta \gamma} \Omega_{{\bf k} \gamma}  f_{{\bf k}}^0 ~ E_{\beta}.
\label{eq:linearintrinsic}
\end{equation}
The first term on the right hand side of the equation above is the usual semiclassical contribution to the conductance that is expressed in terms of the electronic group velocities. This term can provide a finite contribution to the transverse resistance in low-symmetric crystals. However, such contributions are clearly symmetric in the $\alpha \leftrightarrow \beta$ exchange and therefore cannot contribute to the Hall conductance. The second term on the right hand side corresponds instead to the well-known intrinsic contribution to the anomalous Hall conductance controlled by the Berry phases accumulated by the particle motion on the Fermi surface~\cite{Haldane2}. Because the Berry curvature $\Omega_{{\bf k}}$ is an odd function of the momentum when time-reversal is preserved, this Hall conductance vanishes in time-reversal symmetric conditions, in perfect agreement with Onsager relations~\cite{ons31}. 
Let us now instead consider the non-linear response of the current $\propto E^2$. It is simple to show that the non-linear current takes the following form 
\begin{equation} 
j_{\alpha}= \dfrac{e^3 \tau^2}{\hbar^3} \int  \dfrac{d^d k}{(2 \pi)^d} \partial_{k_{\alpha}} \epsilon({\bf k})  \partial_{k_{\beta}}  \partial_{k_{\gamma}} f_{{\bf k}}^0~E_{\beta} E_{\gamma} - \dfrac{e^3 \tau}{\hbar^2}  \int  \dfrac{d^d k}{(2 \pi)^d} \epsilon_{\alpha \beta \delta} \Omega_{{\bf k} \delta}~\partial_{k_{\gamma}} f_{{\bf k}}^0 ~ E_{\beta} E_{\gamma} 
\label{eq:nonlinearintrinsic}
\end{equation}
The first term, which is of entirely semiclassical origin, vanishes in time-reversal symmetric condition since it involves the three index tensor $\partial_{k_{\alpha}} \epsilon({\bf k})  \partial_{k_{\beta}}  \partial_{k_{\gamma}} f_{{\bf k}}^0$ that is odd under time-reversal. 
After integration by parts, the non-vanishing non-linear conductivity tensor can be therefore written as 
\begin{equation}
\chi_{\alpha \beta \gamma} = \dfrac{e^3~\tau}{\hbar^2}   \int  \dfrac{d^d k}{(2 \pi)^d} \epsilon_{\alpha \beta \delta} \left(\partial_{k_{\gamma}} \Omega_{{\bf k} \delta} \right) f_{{\bf k}}^0,
\label{eq:nonlineartensorintrinsic}
\end{equation} 
We therefore have that in time-reversal symmetric conditions the dissipationless Hall non-linear current is regulated by the first moment of the Berry curvature, the BCD~\cite{Sodemann2015}, over the occupied states
\begin{equation}
D_{\gamma \delta}=  \int  \dfrac{d^d k}{(2 \pi)^d}  \left(\partial_{k_{\gamma}} \Omega_{{\bf k} \delta} \right) f_{{\bf k}}^0. 
\label{eq:berrycurvaturedipole}
\end{equation} 
The BCD is subject to the same symmetric constraints of the non-linear Hall conductance introduced in the preceding section. In two-dimensional systems, for instance, the Berry curvature is a pseudoscalar and therefore the BCD is a pseudovector precisely as $\chi_{H \gamma}$. 

In our foregoing discussion, we have considered the dc limit. 
When accounting for an ac driving electric field, non-linear Hall currents yield a current response at twice the driving frequency, with the addition of a rectified current. Specifically for a driving electric field $E_{\alpha} (t)=\textrm{Re} \left\{ \mathcal{E}_{\alpha} e^{i \omega t} \right\}$, with $\mathcal{E} \in {\mathbb C}$, the resulting current 
at twice the frequency  $j_{\alpha}^{2\omega}=\chi_{\alpha \beta \gamma} \mathcal{E}_\beta \mathcal{E}_{\gamma}$ while the rectified current $j_{\alpha}^{0}=\chi_{\alpha \beta \gamma} \mathcal{E}_\beta \mathcal{E}_{\gamma}^{\star}$. The ac non-linear response function takes the following form 
\begin{equation*} 
\chi_{\alpha \beta \gamma} =  \dfrac{e^3~\tau}{2 \hbar^2 \left(1+ i \omega \tau \right)}   \int  \dfrac{d^d k}{(2 \pi)^d} \epsilon_{\alpha \beta \delta} \left(\partial_{k_{\gamma}} \Omega_{{\bf k} \delta} \right) f_{{\bf k}}^0.
\end{equation*} 
It is interesting to note that at frequencies above the width of the Drude peak $\omega \tau \gg 1$ but below the interband transition threshold, the non-linear response function becomes independent of the scattering time and therefore provides a direct measure of a geometric property of the electronic wavefunctions. 

 \subsection{Disorder-induced contributions to the non-linear Hall effect}
 \label{subsec:thqb}
 The BCD does not completely determine the NLHE in a quantum material. Using either a quantum Boltzmann transport approach~\cite{nan19} or a semiclassical Boltzmann theory beyond the constant relaxation time approximation~\cite{xia19,kon19,du19}, it can be shown that other disorder-mediated contributions to the non-linear Hall conductance exist. These ``extrinsic" contributions are the non-linear counterparts of the side-jump and skew-scattering contributions appearing in the linear AHE~\cite{nag10}. In this section, we review the Boltzmann semiclassical transport framework by following Ref.~\cite{du19}, to which we refer the readers interested in further details.  
  
Let us start out by rewriting the collision integral appearing in the kinetic equation Eq.~\ref{eq:boltzmanngen} as
 \begin{equation*} 
{\mathcal I}_{coll} \left\{ f_l \right\}= - \sum_{l^{\prime}} \left(\omega_{l^{\prime} l} f_l - \omega_{l l^{\prime}} f_{l^{\prime}} \right)
 \end{equation*}
 where the label $l$ is a composition a band and momenta indices $\left(n , {\bf k} \right)$ whereas $\sum_l = \sum_n \int d^{d} k / (2 \pi)^d$. In addition, 
  $\omega_{l l^{\prime}}$ is the disorder averaged scattering rate between the Bloch waves with quantum numbers $l$ and $l^{\prime}$. The scattering rate  $\omega_{l l^{\prime}}$ can be related to the T-matrix element 
 \begin{equation*}
  \omega_{l \, l^{\prime}} = \dfrac{2 \pi}{\hbar} \left| T_{l \, l^{\prime}} \right|^2 \delta \left(\epsilon_l - \epsilon_{l^{\prime}} \right). 
  \end{equation*}
  The scattering $T$ matrix is defined by 
  $ T_{l \, l^{\prime}} = \bra{l} {\hat V}_{imp} \ket{\psi_{l^{\prime}}}$ with ${\hat V}_{imp}$ indicating the impurity potential operator whereas $\ket{\psi_l}$ represents the eigenstate of the full Hamiltonian ${\mathcal H}= {\mathcal H}_0+ \hat{V}_{imp}$ that satisfies the Lippman-Schwinger equation 
 \begin{equation*} 
 \ket{\psi_l}=\ket{l} + \dfrac{{\hat V}_{imp}}{\epsilon_l - {\mathcal H}_0 + i \epsilon} \ket{\psi_l}. 
 \end{equation*} 
 For weak disorder one can approximate the scattering state  $\ket{\psi_l}$ by a truncated series in powers of $V_{l l^{\prime}}=\bra{l} {\hat V}_{imp} \ket{l^{\prime}}$ as
 \begin{equation*} 
 \ket{\psi_l}=\ket{l} + \sum_{l^{\prime \prime}} \dfrac{V_{l^{\prime \prime} l}}{\epsilon_l - \epsilon_{l^{\prime \prime}} + i \epsilon} \ket{l^{\prime \prime}} + \ldots 
 \end{equation*}
 Inserting this expression in the expression of the T-matrix element of the disorder potential, we can therefore expand the scattering rate in powers of the disorder strength as
 \begin{equation*}
 \omega_{l l^{\prime}}= \omega_{l l^{\prime}}^{(2)} +  \omega_{l l^{\prime}}^{(3)}+ \omega_{l l^{\prime}}^{(4)} + \ldots
 \end{equation*}
 The scattering rate  $\omega_{l l^{\prime}}^{(2)}= 2 \pi \braket{\left| V_{l l^{\prime}} \right|^2}_{dis} \delta (\epsilon_l - \epsilon_{l^{\prime}}) / \hbar$ is symmetric under the exchange in the state indices $l \leftrightarrow l^{\prime}$. The higher-order corrections contain both a symmetric and an antisymmetric term. The symmetric parts of $\omega_{l l^{\prime}}^{(3,4)}$ can be neglected since they only renormalize the second-order scattering rate $\omega_{l l^{\prime}}^{(2)}$. On the contrary, the antisymmetric contributions to $\omega_{l l^{\prime}}^{(3,4)}$ yield the non-linear skew-scattering and side-jump contributions to the NLHE in time-reversal symmetric conditions. To show this, we first go back to the Boltzmann equation and rewrite explicitly the collision integral using the symmetric and antisymmetric contributions as follows: 
 \begin{equation}
 \dfrac{e}{\hbar} {\bf E} \cdot {\bf \nabla}_{\bf k} f_{l} = - \sum_{l^{\prime}}  \omega_{l l^{\prime}}^{(2)} \left(f_l - f_{l^{\prime}} \right) -  \sum_{l^{\prime}}  \omega_{l^{\prime} l }^{(a)} \left(f_l + f_{l^{\prime}} \right), 
 \label{eq:boltz}
 \end{equation}
 where, now, $\omega_{l^{\prime} l }^{(a)}$ contains the antisymmetric contributions of both  $\omega_{l l^{\prime}}^{(3)}$, and $\omega_{l l^{\prime}}^{(4)}$. The equation above, however, does not account for the microscopic displacement experienced by a wavepacket when scattering from a generic state $l$ to $l^{\prime}$, $\delta {\bf r}_{l l^{\prime}}$. 
Assuming smooth impurity potentials, the gauge-invariant expression for this coordinate shift~\cite{sin06,sin07,sin07b}, usually referred to as side-jump, reads as 
\begin{equation} 
\delta {\bf r}_{l l^{\prime}} = \braket{u_l | i \partial_{\bf k} u_l} -  \braket{u_{l^{\prime}}  | i \partial_{{\bf k}^{\prime}} u_{l^{\prime}}} - \hat{D}_{{\bf k},{\bf k}^{\prime}} \textrm{arg}\left[ \braket{u_l | u_{l^{\prime}}}\right], 
\label{eq:sidejump}
\end{equation}
 where $\textrm{arg}$ is the phase of the complex number and we introduced the operator $ \hat{D}_{{\bf k}~{\bf k}^{\prime}} = \partial_{{\bf k}} + \partial_{{\bf k}^{\prime}}$. 
 In the presence of the external driving electric field, the microscopic displacement $\delta {\bf r}_{l l^{\prime}}$ yields an energy shift $\Delta U_{l l^{\prime}} = - e {\bf E} \cdot \delta {\bf r}_{l l^{\prime}}$, the scattering rate should account for. 
 The symmetric second-order scattering rate, specifically, has to be therefore modified as 
  \begin{eqnarray*}
 \omega_{l l^{\prime}}^{(2)} \rightarrow \omega_{l l^{\prime}}^{(2)}  &=&  \dfrac{2 \pi}{\hbar} \braket{\left| V_{l l^{\prime}} \right|^2}_{dis} \delta \left[ \epsilon_l - \epsilon_{l^{\prime}} - e {\bf E} \cdot \delta {\bf r}_{l l^{\prime}} \right] \\
 & \simeq &  \dfrac{2 \pi}{\hbar} \braket{\left| V_{l l^{\prime}} \right|^2}_{dis} \left[ \delta \left( \epsilon_l - \epsilon_{l^{\prime}} \right) + e {\bf E} \cdot \delta {\bf r}_{l^{\prime} l} \dfrac{\partial}{\partial \epsilon_l}  \delta \left( \epsilon_l - \epsilon_{l^{\prime}} \right) \right]
 \end{eqnarray*}
 where we have used the antisymmetric property [c.f. Eq.~\ref{eq:sidejump}] of the side-jump $\delta {\bf r}_{l l^{\prime}}= - \delta {\bf r}_{l^{\prime} l }$. The Boltzmann equation Eq.~\ref{eq:boltz} is then modified as 
 \begin{equation} 
  \dfrac{e}{\hbar} {\bf E} \cdot {\bf \nabla}_{\bf k} f_{l} = - \sum_{l^{\prime}} \left[\omega_{l l^{\prime}}^{(2)} + e {\bf E} \cdot {\bf O}_{l^{\prime} l} \right] \left( f_l - f_{l^{\prime}} \right) -  \sum_{l^{\prime}}  \omega_{l^{\prime} l }^{(a)} \left(f_l + f_{l^{\prime}} \right), 
  \label{eq:boltzfinal}
  \end{equation}
 where we introduced the quantity 
 \begin{equation*}
 {\bf O}_{l^{\prime} l}= \dfrac{2 \pi}{\hbar}  \braket{\left| V_{l l^{\prime}} \right|^2}_{dis} \delta {\bf r}_{l^{\prime} l} \dfrac{\partial}{\partial \epsilon_l} \delta  \left( \epsilon_l - \epsilon_{l^{\prime}} \right). 
 \end{equation*}
 To proceed further, we decompose the distribution function into three different contributions, {\it i.e.} $f_l = f_l^{int}+ g_l^{sk} +g_l^{adis}$ where the intrinsic distribution function $f_l^{int}=f_l^0 +g_l$, $f_l^0$ being the equilibrium distribution function. We next seek an approximate solution to the Boltzmann equation Eq.~\ref{eq:boltzfinal} by decomposing it into three (time-independent) equations reading
 \begin{eqnarray}
   \dfrac{e}{\hbar} {\bf E} \cdot {\bf \nabla}_{\bf k} f_{l}^{int}&=& -\sum_{l^{\prime}}  \omega_{l l^{\prime}}^{(2)} \left( g_l - g_{l^{\prime}} \right) \nonumber \\ 
      \dfrac{e}{\hbar} {\bf E} \cdot {\bf \nabla}_{\bf k} g_{l}^{adis}&=& -\sum_{l^{\prime}}  \omega_{l l^{\prime}}^{(2)} \left( g_l^{adis} - g_{l^{\prime}}^{adis} \right)  -  e {\bf E} \cdot \sum_{l^{\prime}} {\bf O}_{l^{\prime} l} \left(f_l^{int} - f_{l^{\prime}}^{int} \right) \label{eq:boltzarray} \\ 
      \dfrac{e}{\hbar} {\bf E} \cdot {\bf \nabla}_{\bf k} g_{l}^{sk}&=&  -\sum_{l^{\prime}}  \omega_{l l^{\prime}}^{(2)} \left( g_l^{sk} - g_{l^{\prime}}^{sk} \right) -  \sum_{l^{\prime}}  \omega_{l^{\prime} l }^{(a)} \left(g_l + g_{l^{\prime}} \right) \nonumber
 \end{eqnarray}
As before, we expand the different contributions to the distribution function in a series with each term that vanishes as ${\bf E}^n$. In addition, the term in the collision integral containing the symmetric scattering $\omega^{(2)}_{l l^{\prime}}$ can be solved in the relaxation time approximation. Assuming the latter to be constant for simplicity, we find that the linear and quadratic terms $g_l^{1,2}$  correspond to Eq.~\ref{eq:intrinsicboltz} as expected. Next, we determine the anomalous  $g_l^{adis}$ distribution due to the coordinate shift $\delta {\bf r}_{l l^{\prime}}$. At linear order we have 
\begin{equation*}
0= -\dfrac{g_l^{adis , 1}}{\tau} -  e {\bf E} \cdot \sum_{l^{\prime}}  {\bf O}_{l^{\prime} l} \left(f_l^{0} - f_{l^{\prime}}^{0} \right) .
\end{equation*}
Introducing the velocity contribution due to the accumulation of coordinate shifts after many scattering events~\cite{sin07b,sin05}  ${\bf v}^{sj}_{l} = \sum_{l^{\prime}} \omega^{(2)}_{l l^{\prime}}  \delta {\bf r}_{l^{\prime} l}$, and after straightforward manipulations, the linear anomalous distribution term can be recast as 
\begin{equation}
g_l^{adis , 1}= e \tau {\bf E} \cdot {\bf v}^{sj}_l  \dfrac{\partial f_l^{0} }{\partial \epsilon_l}. 
\label{eq:gadislinear}
\end{equation}
The term of the anomalous distribution quadratic in the driving electric field is instead determined by the equation 
\begin{equation*}
\dfrac{e}{\hbar} {\bf E} \cdot {\bf \nabla}_{\bf k} g_l^{adis, 1} = -\dfrac{g_l^{adis, 2}}{\tau}  - e {\bf E} \cdot \sum_{l^{\prime}} {\bf O}_{l^{\prime}  l} \left( g_l^{1} - g_{l^{\prime}}^{1}  \right).
\end{equation*}
Using Eq.~\ref{eq:gadislinear} and the expression for the linear intrinsic distribution function Eq.~\ref{eq:intrinsicboltz}, we find 
\begin{equation}
g_l^{adis, 2}= - \dfrac{e^2 \tau^2}{\hbar} {\bf E} \cdot {\bf \nabla}_{\bf k} \left( {\bf E} \cdot {\bf v}^{sj}_l \dfrac{\partial f_l^0}{\partial \epsilon_l} \right) + \dfrac{e^2 \tau^2}{\hbar} {\bf E} \cdot \sum_{l^{\prime}} {\bf O}_{l^{\prime} l} \left[ {\bf E} \cdot {\bf \nabla}_{\bf k} f_l^0 - {\bf E} \cdot {\bf \nabla}_{\bf k^{\prime}} f_{l^{\prime}}^0 \right]
\label{eq:gadisquadratic}
\end{equation}
Next, we turn to the skew-scattering contributions to the distribution function. From the third equation in Eq.~\ref{eq:boltzarray}, we find the linear order contribution 
 \begin{equation}
 g_l^{sk, 1} = \dfrac{e \tau^2}{\hbar} \sum_{l^{\prime}} \omega_{l^{\prime} l}^{(a)} \left[ {\bf E} \cdot {\bf \nabla}_{\bf k} f_l^0 + {\bf E} \cdot {\bf \nabla}_{\bf k^{\prime}} f_{l^{\prime}}^0 \right]
 \label{eq:gskewlinear}
 \end{equation}
 Using the equation above, we can also determine the first non-linear contribution due to skew-scattering to the distribution function. It reads: 
\begin{equation}
 g_l^{sk, 2} = - \dfrac{e^2 \tau^3}{\hbar^2} \left\{ {\bf E} \cdot {\bf \nabla}_{\bf k} \sum_{l^{\prime}} \omega_{l^{\prime} l}^{(a)} \left[ {\bf E} \cdot {\bf \nabla}_{\bf k} f_l^0 + {\bf E} \cdot {\bf \nabla}_{\bf k^{\prime}} f_{l^{\prime}}^0 \right] + \sum_{l^{\prime}} \omega_{l^{\prime} l}^{(a)} \left[ {\bf E} \cdot {\bf \nabla}_{\bf k} \left( {\bf E} \cdot {\bf \nabla}_{\bf k} f_l^{0} \right) +  {\bf E} \cdot {\bf \nabla}_{\bf k} \left( {\bf E} \cdot {\bf \nabla}_{\bf k} f_l^{0} \right) \right] \right\}. 
 \label{eq:gskewquadratic}
 \end{equation}
 
 Having in our hands the anomalous and the skew-scattering linear and non-linear distribution functions, we can now evaluate the response current to a driving electric field ${\bf j}=e \sum_l \dot{\bf r} f_l$. In doing so, we note that the side-jump velocity introduced above explicitly enters into the semiclassical equation of motion ~\cite{sin05,sin06,sin07,sin07b,nag10,xia10,du19}
 \begin{equation*}
 \dot{\bf r}_l= \dfrac{1}{\hbar} {\bf \nabla}_{\bf k} \epsilon_l  + \dfrac{e}{\hbar} {\bf E} \cross {\bf \Omega}_l + {\bf v}_l^{sj}  
 \end{equation*}
This implies that in the linear response regime there are three disorder-mediated contributions beyond the semiclassical and anomalous Hall conductance of Eq.~\ref{eq:linearintrinsic}. 
There is a first contribution due to the anomalous side-jump distribution that yields a conductivity 
\begin{equation}
\sigma^{sj ,1}_{\alpha \beta} = \dfrac{e^2 \tau}{\hbar} \sum_l v_{l \, \beta} ^{sj} ~{\bf \nabla}_{{k}_{\alpha}} \epsilon_l ~ \dfrac{\partial f_l^0}{\partial \epsilon_l}. 
\label{eq:sidejump1linear}
\end{equation}
In addition, the side-jump velocity appearing in the equation of motion contributes with an additional conductivity 
\begin{equation}
\sigma^{sj , 2}_{\alpha \beta} = - \dfrac{e^2 \tau}{\hbar} \sum_{l} v_{l \, \alpha}^{sj} ~ {\bf \nabla}_{{k}_{\beta}} \epsilon_l ~ \dfrac{\partial f_l^0}{\partial \epsilon_l}. 
\label{eq:sidejump2linear}
\end{equation}
We point out that we have neglected terms $\propto {\bf v}_l^{sj} g^{(adis, sk), 1}$ since they are of higher-order power in the scattering rate $\omega$. 
Finally, the skew-scattering contribution to the distribution function leads to the conductivity 
\begin{equation}
\sigma^{sk}_{\alpha \beta} = - \dfrac{e^2 \tau^2}{\hbar^2} \sum_{l l^{\prime}} \omega_{l l^{\prime}}^{(a)} \left[{\bf \nabla}_{k_{\alpha}} \epsilon_l ~ {\bf \nabla}_{k_{\beta}} \epsilon_l - {\bf \nabla}_{k_{\alpha}} \epsilon_{l^{\prime}} ~ {\bf \nabla}_{k_{\beta}} \epsilon_l \right] \dfrac{\partial f_l^0}{\partial \epsilon_l}
\label{eq:skewlinear}
\end{equation}
In systems with broken time-reversal symmetry,  Eqs.~\ref{eq:sidejump1linear},\ref{eq:sidejump2linear},\ref{eq:skewlinear} provide the side-jump and skew-scattering contributions to the anomalous Hall conductance. We note that Eq.~\ref{eq:skewlinear} can be split into two different contributions corresponding to the antisymmetric scattering rates $\omega^{(3)}$ and $\omega^{(4)}$~\cite{nag10}. 
In time-reversal symmetric conditions these extrinsic contributions to the Hall conductance are forced to vanish, precisely as the intrinsic Berry-curvature mediated anomalous Hall conductance. 
This is because the coordinate shift $\delta {\bf r}_{l^{\prime} l}$, and consequently the side-jump velocity ${\bf v}_l^{ s j}$, are even under time-reversal. Since the conventional group velocity $ \propto {\bf \nabla}_{\bf k}$ is instead odd under time-reversal, one finds that 
side-jump conductivities $\sigma^{sj, (1,2)}$ are forced to vanish. The same holds true for the skew-scattering contribution since the antisymmetric scattering rate is odd under time-reversal.

 We now evaluate the non-linear disorder-induced conductivity. Using the aforementioned symmetry constraints, it is easy to show that the terms quadratic in the driving electric field where the anomalous velocity ${\bf \Omega}$ is directly coupled to the anomalous distribution $g_l^{adis , 1}$ and the skew-scattering distribution $g_l^{sk , 1}$ are forced to vanish by time-reversal. 
As a result, we have that the only Berry curvature mediated term corresponds to the dipole of Eq.~\ref{eq:nonlineartensorintrinsic}. 
As for the linear conductivity, there are three disorder-induced non-linear terms. 
The first contribution comes from the second-order anomalous distribution function $g_l^{adis, 2}$. From Eq.~\ref{eq:gadisquadratic} and after simple manipulations it can be recast as 
\begin{equation}
\chi_{\alpha \beta \gamma}^{sj , 1}=  \dfrac{e^3 \tau^2}{\hbar^2} \sum_l \left[\left({\bf \nabla}_{k_{\alpha}} v_{l \beta}^{sj} + {\mathcal M}_{l l^{\prime}}^{\alpha \beta} \right) {\bf \nabla}_{k_{\gamma}} \epsilon_l +  \left({\bf \nabla}_{k_{\beta}} {\bf \nabla}_{k_{\alpha}} \epsilon_l \right) v_{l \, \gamma}^{sj} \right]~\dfrac{\partial f_l^0}{\partial \epsilon_l} ,
\label{eq:sidejump1quadratic}
\end{equation}
where we introduced the tensor 
\begin{equation}
{\mathcal M}_{l l^{\prime}}^{\alpha \beta} = \dfrac{ 2 \pi}{\hbar} \sum_{l^{\prime}} \left[{\bf \nabla}_{k_{\alpha}}  \left( \braket{\left| V_{l l^{\prime}} \right|^2}_{dis} \delta {\bf r}_{l l^{\prime}} \right)-  {\bf \nabla}_{k^{\prime}_{\alpha}}  \left(  \braket{\left| V_{l l^{\prime}} \right|^2}_{dis} \delta {\bf r}_{l^{\prime} l} \right) \right] \delta \left(\epsilon_l - \epsilon_{l^{\prime}} \right)
\label{eq:sidejumptensor}
\end{equation}
The contribution due to the side-jump velocity can be instead put as 
\begin{equation}
\chi_{\alpha \beta \gamma}^{sj , 2} = \dfrac{e^3 \tau^2}{\hbar^2} \sum_l v_{l \, \alpha}^{sj} {\bf \nabla}_{k_{\beta}} {\bf \nabla}_{k_{\gamma}} f_l^0. 
\label{eq:sidejump2quadratic}
\end{equation}
Finally, the skew scattering contribution to the non-linear distribution function yields the additional conductivity 
\begin{equation}
\chi_{\alpha \beta \gamma}^{sk}= \dfrac{e^3 \tau^3}{\hbar^3} \sum_{l l^{\prime}} \omega_{l l^{\prime}}^{(a)} \left[\left( {\bf \nabla}_{k_{\alpha}} \epsilon_l - {\bf \nabla}_{k^{\prime}_{\alpha}} \epsilon_{l^{\prime}} \right) {\bf \nabla}_{k_{\beta}} {\bf \nabla} _{k_{\gamma}} f_l^0 - \left( {\bf \nabla}_{k_{\beta}} {\bf \nabla}_{k_{\alpha}} \epsilon_l -  {\bf \nabla}_{k^{\prime}_{\beta}} {\bf \nabla}_{k^{\prime}_{\alpha}} \epsilon_{l^{\prime}}  \right) {\bf \nabla}_{k_\gamma} f_l^0 \right]
\label{eq:skewquadratic}
\end{equation}
Eqs.~\ref{eq:sidejump1quadratic},\ref{eq:sidejumptensor},\ref{eq:sidejump2quadratic},\ref{eq:skewquadratic}  provide the disorder-induced contributions to the non-linear Hall effect in the dc limit and assuming a constant relaxation time. 
As mentioned above, these results can be easily generalized assuming a driving a.c. electric field and a general relaxation time (see Ref.~\cite{du19} for further details). 
In closing this section, we point out that disorder-induced contributions to the non-linear Hall conductance have a dependence in the impurity concentration $n_i$ given by  $\chi^{sj, sk} \propto (n_i V_0^2)^{-1}$, with $V_0^2= \braket{V_i^2}_{dis}$ and $V_i$ the random disorder strength. This also implies that precisely as the BCD contribution Eq.~\ref{eq:nonlineartensorintrinsic}, the disorder-mediated corrections to the non-linear Hall conductance grow linear with the relaxation time. 
Nevertheless, the different contributions to the non-linear Hall conductance can be distinguished in the experimental realm by using the scaling between the non-linear Hall signal and the conventional longitudinal (linear) resistivity, in strict analogy with the scaling used in the AHE~\cite{tia09,hou15}.

\section{Strongly spin-orbit coupled 2D materials }
\label{sec:2dsoc}
\subsection{Tilted massive Dirac cones} 
\begin{figure}
  \includegraphics[width=\linewidth]{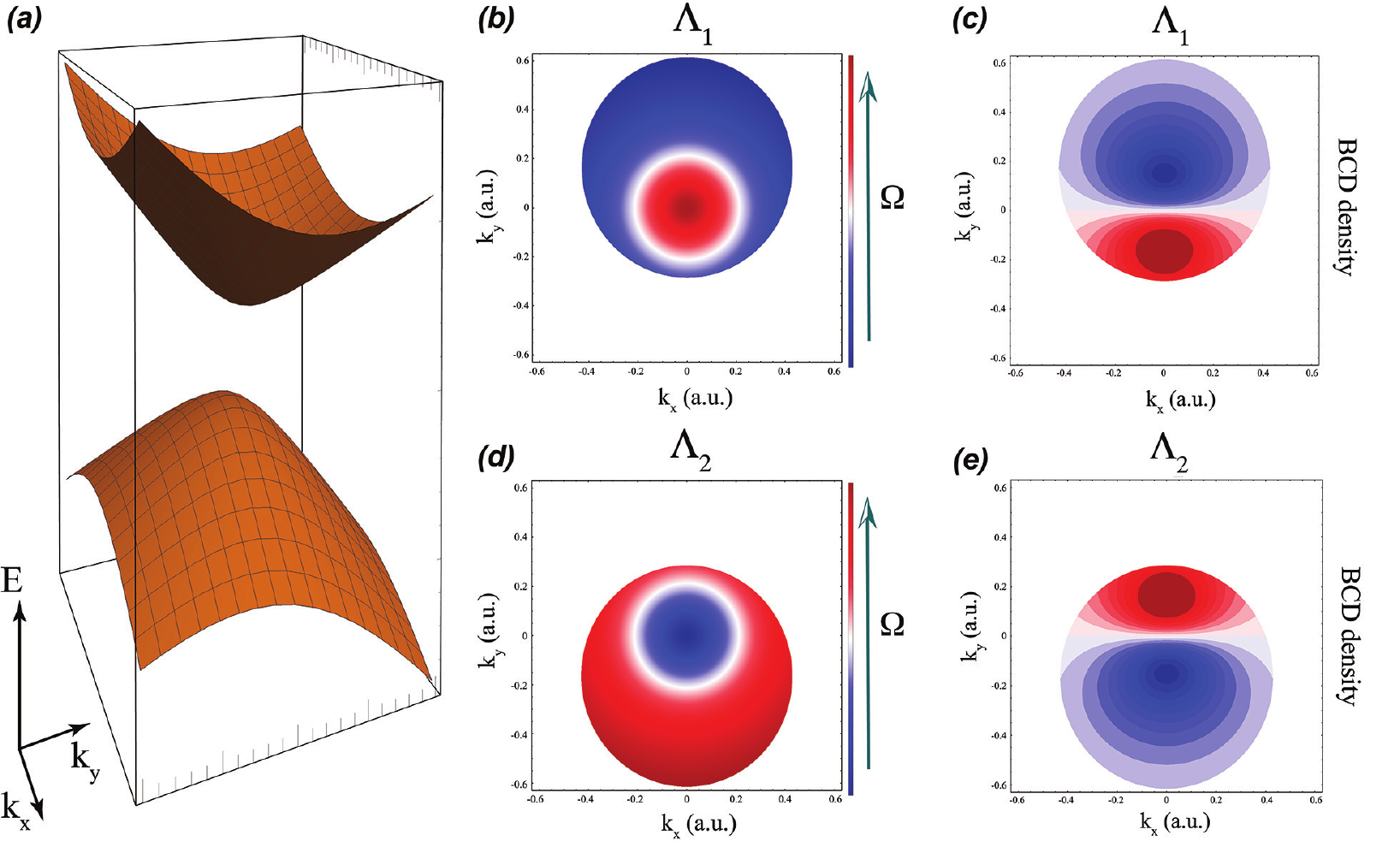}
  \caption{{\bf (a)} Energy dispersion of a tilted massive Dirac cone with isotropic Fermi velocity, {\it i.e.} $v_x=v_y=v$. We have chosen the tilt parameter $a/v_F=0.1$ and a Dirac mass $m/v_F=0.3$. {\bf (b),(c)} Corresponding density plot of the Berry curvature and of the Berry curvature dipole density considering a Fermi energy $\epsilon_F=v k_0 /2$ with $k_0$ a characteristic momentum. {\bf (d),(e)} Same for the tilted massive Dirac cone appearing in the time-reversal related valley.}
  \label{fig:tiltedcones}
\end{figure}

Beside the general symmetry constraints discussed in Sec.~\ref{subsec:thsymm}, the Berry curvature-mediated contribution to the non-linear Hall conductance is subject to other point-group symmetry restrictions, which are of primary importance in identifying materials that possess substantial BCD. Let us first consider specifically the role played by rotational symmetries.  
As long as time-reversal symmetry is preserved, all systems with an evenfold rotation symmetry ${\mathcal C}_{n}$ (with $n=2,4,6$) cannot have a finite BCD. 
This is because the composed symmetry ${\mathcal C}_2 \Theta$, $\Theta$ being the time-reversal operator, is an antiunitary symmetry that squares to one. Moreover this symmetry acts locally in momentum space since it brings the two-dimensional momentum ${\bf k}$ back to itself. These two properties imply that  the Berry curvature is forced to vanish for all momenta in the two-dimensional BZ~\cite{bat21}. 
Consequently, a non-vanishing BCD can only appear in two-dimensional crystals where  there is either a threefold rotation symmetry or all rotation symmetries are broken. 
For the latter and as discussed in Sec.~\ref{subsec:thsymm}, a single mirror line can still exists, in which case the BCD is forced to be orthogonal to the mirror line. 
In the former case, instead, the BCD will not be pinned to any specific direction. 
The time-reversal symmetric NLHE in a ${\mathcal C}_3$-symmetric setting has been not proposed or experimentally realized so far. 
Therefore, in the remainder we will limit ourselves to discuss systems with a single mirror line symmetry, {\it i.e.} with point group symmetry ${\mathcal C}_{s}$. 
In order to have a sizable Berry curvature, the low-energy electronic properties of such systems must be described by massive Dirac cones. 
Furthermore, the absence of rotational symmetries implies that there is no fermion multiplication theorem~\cite{fan19}, and the  minimum number of massive Dirac cones allowed by symmetry is two~\cite{nie81}.  
To preserve the mirror symmetry these two massive Dirac cones will be located specularly with respect to the mirror-symmetric line of the two-dimensional BZ. 
To make things concrete, consider for instance the reflection symmetry to map a point with coordinates $\left( x , y \right)$ to $\left( x , - y \right)$. The massive Dirac cones will then be centered around two valleys $\Lambda_{1,2}= \left\{{\bar k}_x, \pm {\bar k}_y \right\}$. 
The effective Hamiltonian close to these valleys can be derived by accounting for all symmetry-allowed terms in a ${\bf k} \cdot {\bf p}$ expansion. 
Time-reversal symmetry gives the constraint on the effective Hamiltonian $\Theta^{-1} {\mathcal H}_{eff} \left(k_x, k_y \right) \Theta =  {\mathcal H}_{eff} \left(-k_x, -k_y \right)$. Here $\Theta$ is the time-reversal operator that is represented as $\Theta=i \sigma_y \tau_x {\mathcal K}$ where the Pauli matrix vectors ${\boldsymbol \sigma}$ and ${\boldsymbol \tau}$   act in spin and valley space respectively, whereas ${\mathcal K}$ is the complex conjugation. 
Similarly, the mirror symmetry gives the constraint ${\mathcal M}_y^{-1}  {\mathcal H}_{eff} \left(k_x, k_y \right) {\mathcal M}_y=  {\mathcal H}_{eff} \left(k_x, -k_y \right)$. The mirror symmetry operator takes the form ${\mathcal M}_y= - i \sigma_y \tau_x$ since it exchanges the valleys, as time-reversal does, and acts in spin space as $\exp{- i \pi \sigma_y / 2}$. 
Retaining terms up to linear order in momentum ${\bf k}$ and neglecting intervalley mixing terms -- this condition amounts to keep only symmetry allowed terms $\propto \tau_0, \tau_z$ -- the low-energy continuum Hamiltonian gets the following form: 
\begin{equation}
{\mathcal H}_{eff} = a \tau_z \sigma_0 k_y + \left(v_x k_x \sigma_y - v_y k_y \sigma_x \right) \otimes \tau_0 + m \tau_z \sigma_z. 
\label{eq:Diractilted}
\end{equation}
In the equation above we have neglected a term of the form $\alpha \sigma_x \tau_z$ since this term shifts in a time-reversal symmetric manner the two valleys $\Lambda_{1,2}$, and can be thus reabsorbed in the effective Hamiltonian by a proper redefinition of the momentum $k_y$. 
Eq.~\ref{eq:Diractilted} corresponds to the low-energy theory of a massive Dirac cone except for the term proportional to $a$, which produces a tilt as shown in Fig.~\ref{fig:tiltedcones}(a). 

\begin{figure}
\begin{center}
  \includegraphics[width=.65\linewidth]{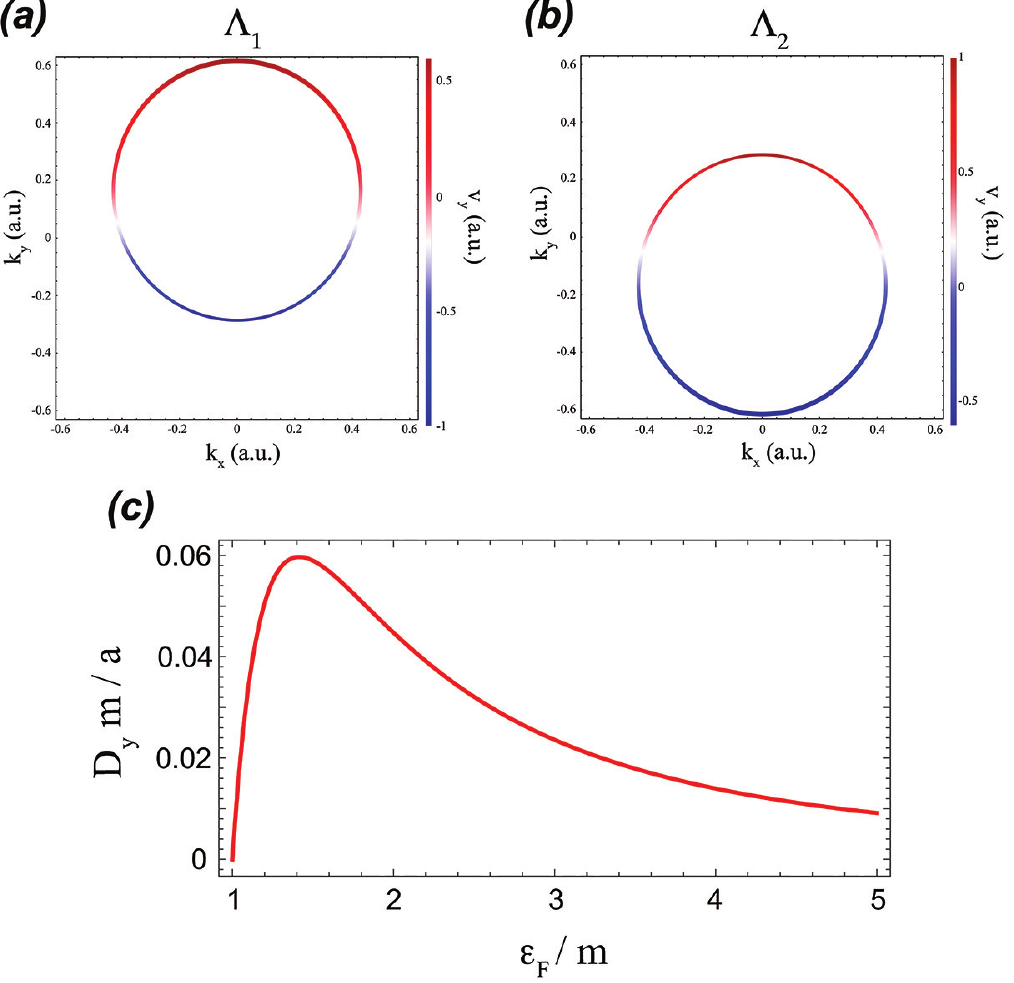}
\end{center} 
  \caption{{\bf (a),(b)} Contour plot of the group velocity $|v_y|$ in a pair of tilted massive Dirac cones. The parameter set is same as in Fig.~\ref{fig:tiltedcones}. {\bf (c)} Behavior of the Berry curvature dipole in a systems with a pair of tilted massive Dirac cones as regulated by Eq.~\ref{eq:dipoletiltedanalytical}.}
  \label{fig:tiltedvelocity}
\end{figure}

The presence of this tilt term is key for the occurrence of a finite BCD~\cite{nan19,Sodemann2015,du19}, as we now discuss. It is straightforward to show that the Berry curvature is independent of the tilt strength and can be written as   
\begin{equation} 
\Omega_{{\bf k} z}^\tau= \pm  \,  \dfrac{\tau v_x v_y m}{2~\left( m^2 + v_x k_x^2 + v_y k_y^2 \right)^{3/2}}, 
\end{equation} 
where the $\pm$ sign refers to the conduction and valence bands respectively, and $\tau=\pm 1$ for the two valleys $\Lambda_{1,2}$. 
In each valley the Berry curvature is an even function of the momentum ${\bf k}$ measured relatively to the Dirac points. Therefore, the BCD density $\partial_{k_y} \Omega_{{\bf k} z}^\tau$ is an odd function of the momentum. Consequently, the contribution to the BCD coming from each valley is identically zero in the absence of a tilt term. 
The situation is different for $a \neq 0$. In this case the Fermi lines become elliptical, and therefore the contribution to the BCD in each valley is different from zero  [c.f. Fig.~\ref{fig:tiltedcones}(b-e)]. In addition, the contribution from the two different valleys is same since the elliptical distortion of the Fermi lines is opposite in the two valleys. As a final result, we therefore have that the BCD density is directly  proportional to the tilt parameter $a$. 
Another way to understand the tilt-driven occurrent of a finite BCD is to transform $D_y$ as a line integral over the Fermi line of the system as  
\begin{equation}
D_y= \sum_{\tau} \int \dfrac{d^2 k}{(2 \pi)^2} \Omega_{{\bf k} z}^\tau {\bf v}_{\bf k}^y \times \delta\left(\epsilon-\epsilon_F \right). 
\label{eq:dipoleline}
\end{equation}
In the absence of a tilt term, this integral vanishes in each valley since the Berry curvature is constant on the Fermi lines whereas the group velocity in the ${\hat y}$ direction is equal but opposite on the opposite sides of the Fermi surface. Tilting the Dirac cones allows for a mismatch between left and right movers in each valley [see Fig.~\ref{fig:tiltedvelocity}(a),(b)] leading to a finite contribution to the BCD. Furthermore, the contribution to the dipole coming from the two valleys is equal since the opposite mismatch between left and right movers is compensated by the opposite values of the Berry curvature.  
Using Eq.~\ref{eq:dipoleline} it is possible to provide an analytical expression for the BCD in systems with tilted massive Dirac cones. Consider for simplicity the case in which $v_x=v_y=v$, and assume the tilt parameter $a$ is small as compared to the Fermi velocity $v$. Expanding the Dirac delta as
$\delta\left(\epsilon-\epsilon_F \right) = \delta\left(\epsilon^0-\epsilon_F \right) + a \tau k_y \partial_{\epsilon^0} \delta\left(\epsilon^0-\epsilon_F \right)$, with $\epsilon^0$ indicating the energy of the untilted massive Dirac cone, the BCD can be found to be
\begin{equation}
D_y = \dfrac{3 a}{4 \pi}~\dfrac{m}{\epsilon_F^4}~\left[\epsilon_F^2 - m^2 \right]
\label{eq:dipoletiltedanalytical}
\end{equation}
where we considered the Fermi energy to be in the conduction band. It is important to note that as a function of the chemical potential, the BCD displays a characteristic non-monotonous behavior~\cite{Sodemann2015} [see Fig.~\ref{fig:tiltedvelocity}(c)]. A similar feature is also found when explicitly computing the disordered-averaged contributions to the NLHE discussed in the previous section~\cite{nan19,du19}. The overall contribution does not cancel out and is still finite. 

\subsection{Surface states of topological crystalline insulators}
\begin{figure}
\begin{center}
  \includegraphics[width=.75\linewidth]{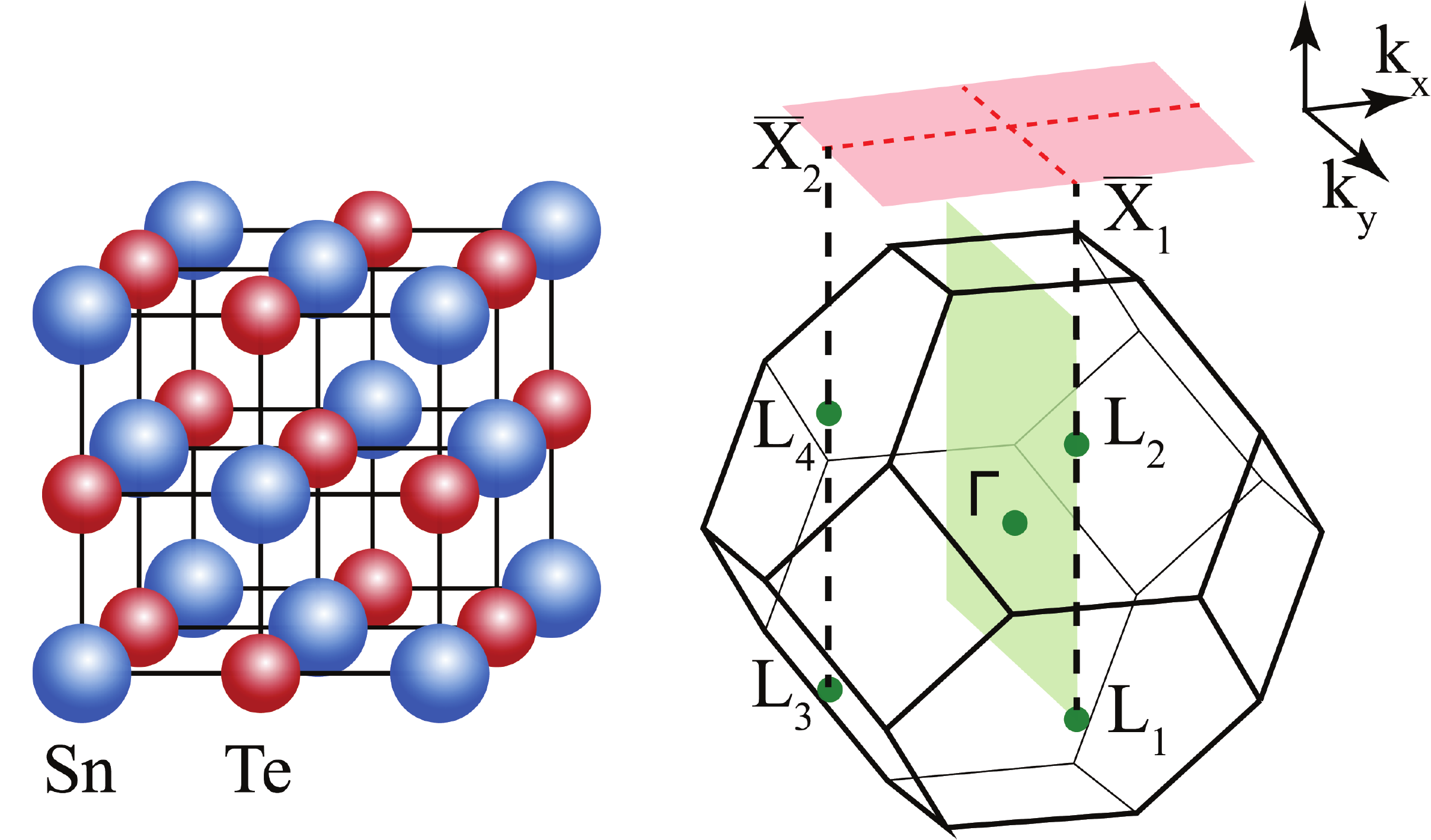}
\end{center} 
  \caption{Left panel: High-temperature rocksalt crystal structure of SnTe. Right panel: Corresponding Brillouin zone with the high-symmetry points $L$, the Brillouin zone center $\Gamma$ and the mirror symmetric plane. We also show the $(001)$ surface BZ projection with highlighted high-symmetry lines.}
  \label{fig:tci}
\end{figure}

The surface states of the topological crystalline insulator SnTe~\cite{hsi12,tan12,and15} represent a prime example of tilted massive Dirac cones. This makes SnTe a paradigmatic material platform for the realization of the quantum NLHE in time-reversal symmetric conditions.  
Generally speaking, the group-IV tellurides have a high-temperature rocksalt crystal structure with a face-centered cubic Brillouin zone (BZ) [c.f. Fig.~\ref{fig:tci}]. The BZ is bounded by six square faces and eight hexagonal faces~\cite{hsi12}. The centers of the latter, commonly denoted by $L$, represent the equivalent high-symmetry points of the BZ where the fundamental gap of the SnTe or PbTe is located. The band structure in the immediate vicinity of the four equivalent $L$ points can be captured using a four-band  ${\bf k} \cdot {\bf p}$ model~\cite{mit66} that span spin space and the $p$-orbital space of the anion (Te) and cation (Pb,Sn). In PbTe, the valence band is derived from the $p$-orbital of the anion while the conduction band from the cation Pb. On the contrary, SnTe has an inverted band ordering with the valence band derived from the cation Sn and the conduction band from Te. The inverted ordering of the bands of SnTe as compared to PbTe establishes the former as a topological crystalline insulator. The bulk crystalline topology of the material comes about the non-trivial mirror Chern number 
associated to the  $\Gamma L_1 L_2$ plane of the BZ [see Fig.~\ref{fig:tci}], which is left invariant by the $\left(1 1 0 \right)$ mirror symmetry of the rocksalt crystal structure. The non-trivial value of the mirror Chern number then implies the presence of counterpropagating midgap modes on the surface projections of the mirror invariant plane~\cite{hsi12,lau18}. This, for instance, applies to the $\bar{\Gamma}-\bar{X}_1$ line of the $\left( 0 0 1 \right)$ surface BZ. Since the mirror Chern number associated to the  $\Gamma L_1 L_2$ plane $n_{\mathcal M}=2$, there will be two pairs of counterpropagating edge modes along the $\bar{\Gamma}-\bar{X}_1$ line. In addition, 
since the left and right edge modes belong to the different $\pm i$ mirror sectors, their crossings cannot be gapped. On the contrary, away from the $\bar{\Gamma} - \bar{X}_1$ line the in-gap edge modes can be gapped since they are not protected by mirror symmetry. The crossings on the $\bar{\Gamma}-\bar{X}_1$ line of the $\left(0 0 1 \right)$ surface BZ are the Dirac points of two topologically protected surface Dirac cones. By rotational symmetry, such surface Dirac cones also appear along the $\bar{\Gamma}-\bar{X}_2$ line. 
One can derive the electronic characteristics of these four surface Dirac cones by using a two-dimensional ${\bf k} \cdot {\bf p}$ theory close to the $\bar{X}_{1,2}$ points of the surface BZ~\cite{liu13,ser14}. 
It can be found by simply noticing that the ${\bar X}_{1,2}$ points of the surface BZ are invariant under a twofold rotation symmetry ${\mathcal C}_2$, and the two mirror line symmetries ${\mathcal M}_{x,y}$ -- from here onwards we denote with $\hat{x}$ the $\left(1 1 0 \right)$ direction perpendicular to the mirror plane. As in our former analysis, these symmetry operations, together with time-reversal symmetry, constrain the allowed terms in the ${\bf k} \cdot {\bf p}$ expansion for the surface Dirac cones. 
Furthermore, the representation of the four symmetries can be derived by noticing that at the ${\bar X}_{1,2}$ points of the $\left(0 0 1 \right)$ surface BZ two out of the four equivalent $L$ points of the BZ are projected [see Fig.~\ref{fig:tci}]. Imagine now that instead of considering an atomically sharp interface between the vacuum and the material bulk, one would consider a smooth interface separating SnTe from PbTe. Then, one would expect Dirac-like domain wall states, each of which coming about from the change in the mass gap in one specific $L$ point of the BZ. At the $\bar{X}_{1,2}$ one would then expect two different flavours of Dirac-like states. Consequently, the effective surface ${\bf k} \cdot {\bf p}$ theory should account for such a flavor degree of freedom beside the internal spin degree of freedom. This holds true even when the smooth interface is substituted by an atomically sharp interface. With this in mind, we can proceed to find the representations of the symmetry operations that constraint the surface low-energy theory. Let us explicitly consider the low-energy theory for the surface states projected onto the ${\bar X}_1$ point of the surface BZ. The theory close to the $\bar{X}_2$ point can be obtained following the same strategy. 

The internal time-reversal symmetry does not act on the flavor degree of freedom. Therefore it can be represented, as usual, by $\Theta= i \sigma_y {\mathcal K}$ where ${\mathcal K}$ is the complex conjugation and the Pauli vector $\boldsymbol{\sigma}$ acts in spin space. The twofold rotation operator ${\mathcal C}_{2}$ interchanges the $L_{1,2}$ valleys while it can be represented in spin space by $\exp{- i \pi \sigma_z / 2}$. As a result, the twofold rotation operator is represented by ${\mathcal C}_2 = - i \, \tau_x \, \sigma_z$. Similarly to the twofold rotation operator ${\mathcal C}_2$, also the mirror symmetry ${\mathcal M}_y$ interchanges the  $L_{1}$ point with $L_2$. Hence, this mirror symmetry can be represented as ${\mathcal M}_y = - i \sigma_y \tau_x$. On the contrary, the mirror symmetry ${\mathcal M}_x= - i \sigma_x \tau_0$ as it does not interchanges the two projected $L$ points. 
With this, we can write the surface ${\bf k} \cdot {\bf p}$ theory  away from the $\bar{X}_{1}$ point of the surface BZ. Specifically, by retaining flavor mixing terms up to zeroth order in momentum the low-energy theory can be found to be
\begin{equation}
{\mathcal H}_{{\bf k} \cdot {\bf p}} = m \tau_x \sigma_0 + \delta \tau_y \sigma_x + \left( v_{x} k_x \sigma_y - v_y k_y \sigma_x \right) \otimes \tau_0. 
\label{eq:kptci}
\end{equation}
As expected, the surface states are gapped specifically on the $k_y=0$ line, {\it i.e.} along the ${\bar X}_{1} - {\bar M}$ direction of the surface BZ. On the contrary the ${\mathcal M}_x$ symmetry guarantees the presence of two zero-energy Dirac points for $k_y = \pm v_y / \sqrt{m^2 + \delta^2}$ on the $k_x=0$ line. In perfect agreement with the foregoing general symmetry analysis, the in-gap surface states realize two Dirac cones located at $\Lambda_{1,2}$ with an effective Hamiltonian 
\begin{equation}
{\mathcal H}_{Dirac} = \widetilde{v}_x  \delta k_x s_y - v_y \delta k_y s_x 
\label{eq:diracsstci}
\end{equation}
where we introduced the momentum $\delta {\bf k} = {\bf k} -{\bf \Lambda}_{1,2}$ and the renormalized velocity $\widetilde{v}_x = v_x \delta / \sqrt{m^2 + \delta^2}$. 
As mentioned above, this form of the surface states for SnTe has been found by neglecting flavour mixing term $ \propto \tau_{x,y,z}$ linear in momentum. 
If the latter were to be considered, the following perturbation should be added to the original surface ${\bf k} \cdot {\bf p}$ Hamiltonian: 
\begin{equation}
\Delta {\mathcal H}_{{\bf k} \cdot {\bf p}} = \left( \alpha_x  k_x \sigma_y  - \alpha_y k_y \sigma_x \right) \otimes \tau_x + \beta k_y \sigma_0 \tau_y  + \gamma k_x \sigma_z \tau_z 
\end{equation}
Projecting this terms in the low-energy Hamiltonian reveals that the surface Dirac dispersion of Eq.~\ref{eq:diracsstci} is modified in two different ways. 
First, it leads to an additional renormalization of the Fermi velocity $\widetilde{v}_x$. Second, it yields a tilt of the surface Dirac cones. The corresponding low-energy theory indeed reads
\begin{equation}
{\mathcal H}_{Dirac} = a \chi \delta k_y s_0 +  \widetilde{v}_x  \delta k_x s_y - v_y \delta k_y s_x,
\end{equation}
where the tilt parameter $a$  is directly proportional to the ${\bf k} \cdot {\bf p}$ parameters $\alpha_y, \beta$ whereas $\chi=\pm 1$ distinguishes the $\Lambda_{1,2}$ valleys.  

The absence of a surface energy gap does not allow for a finite Berry curvature. The latter, however, naturally arises~\cite{oka13,woj15} when considering a ferroelectric distortion whereby the Sn and Te atoms are displaced along different directions. SnTe is known to undergo a structural transition at low temperatures~\cite{rab85,sal10,mur17} that involves precisely such a ferroelectric distortion. Besides yielding a finite ferroelectric polarization, this structural distorsion breaks completely the rotational symmetry~\cite{lau19}. Moreover if the displacement vector is along the ${\hat x}$ direction, the structural distortion breaks  the ${\mathcal M}_x$ mirror symmetry, whereas a displacement in the orthogonal direction leads to a loss of the ${\mathcal M}_y$ reflection symmetry. 
Using the preceding symmetry analysis, it can be shown that breaking the ${\mathcal M}_x$ symmetry changes the crossings of the states along the $\Gamma-\bar{X}_1$ line into avoided level crossings, and ultimately the Dirac cones centered at $\Lambda_{1,2}$ acquire a mass gap. Breaking instead the ${\mathcal M}_x$ symmetry preserves massless Dirac fermions close to the ${\bar X}_1$ point of the surface BZ~\cite{ser14}. The situation is of course reversed when considering the surface states located close to the ${\bar X}_2$ point of the surface BZ. All in all, we therefore have that a ferroelectric distortion with the displacement vector oriented along a principal crystallographic direction of the surface BZ will result in the gapping of two out of the four total  $\left(0 0 1 \right)$ surface Dirac cones in SnTe. 

To concretely show how a mass gap for the surface Dirac cones at $\Lambda_{1,2}$ appears when the ${\mathcal M}_x$ reflection symmetry is broken, it is sufficient to notice that in this situation an additional symmetry-allowed term has to be included in the ${\bf k} \cdot {\bf p}$ Hamiltonian of Eq.~\ref{eq:kptci}. It reads 
\begin{equation}
{\mathcal H}_{{\bf k} \cdot {\bf p}}^{FE} = w_x \tau_y \sigma_z. 
\end{equation} 
Projecting this term onto the low-energy surface state Hamiltonian, we find that the ferroelectric distortion yields two opposite masses for the two surface Dirac fermions centered at $\Lambda_{1,2}$. Therefore, the full Hamiltonian can be recast as 
\begin{equation}
{\mathcal H}_{Dirac} = a \chi  \delta k_y s_0 +  \widetilde{v}_x  \delta k_x s_y - v_y  \delta k_y s_x + \beta \chi s_z,  
\end{equation}
where $\chi=\pm 1$ distinguishes the $\Lambda_{1,2}$ valleys. 
We have therefore reached the massive tilted Dirac fermion model introduced in the preceding section and characterized by a non vanishing BCD. 

\subsection{Transition metal dichalcogenide monolayers}
\begin{figure}
\begin{center}
\includegraphics[width=.65\linewidth]{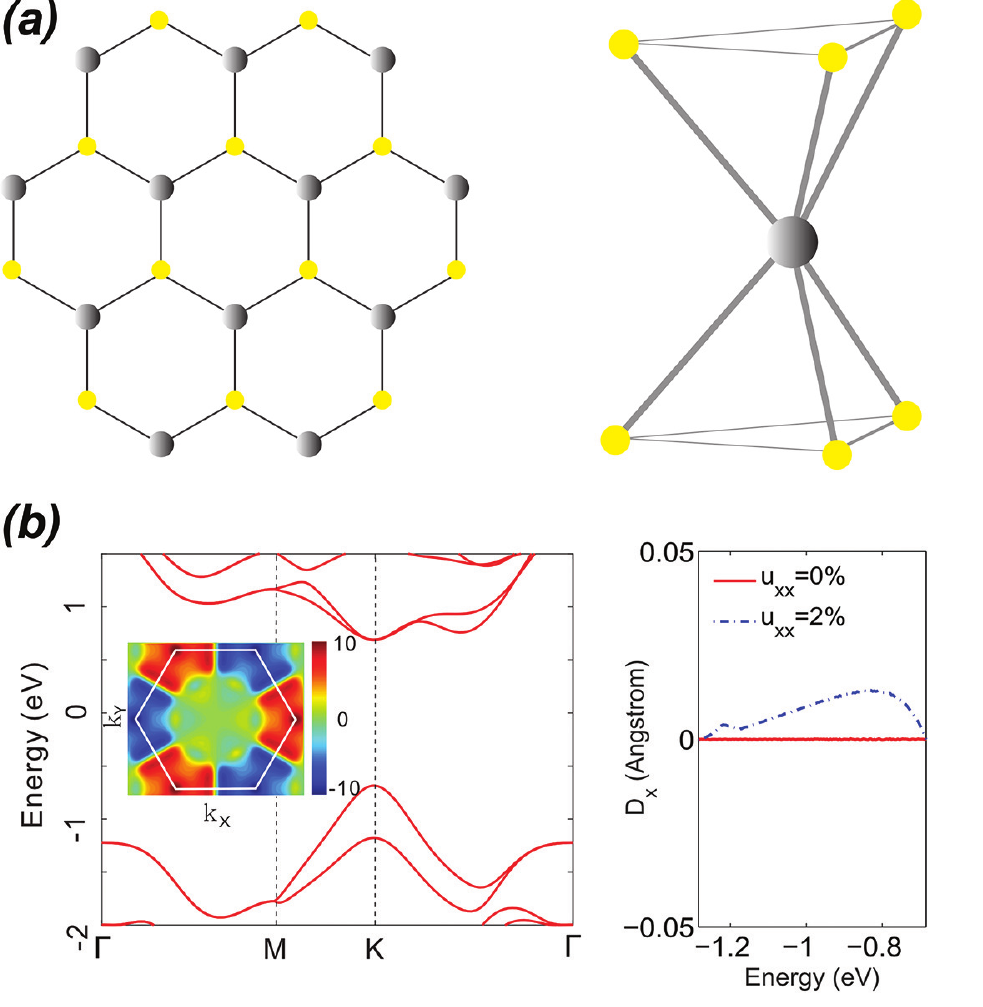}
\end{center} 
\caption{{\bf (a)} Top view of monolayer MX$_2$ in their 1H hexagonal structure.  The grey sphere is the transition metal M while the yellow spheres are the M chalcogen atoms. On the right we also show a sketch of the trigonal prismatic coordination structure. {\bf (b)} Berry curvature dipole of monolayer hexagonal WSe$_2$ with uniaxial strain. Panel {\bf (b)} is reproduced with permission.\textsuperscript{[Ref.~\cite{you18}]} Copyright 2018, American Physical Society.}
\label{fig:TMD1H}
\end{figure}

Having established the occurrence of a finite BCD from the topologically protected surface states of SnTe, we next show the occurrence of this phenomenon also in two-dimensional crystalline materials characterized by a strong spin-orbit coupling. Due to the presence of massive Dirac cones, transition metal dichalcogenides (TMDs) in their so-called $1H$ form [see Fig.~\ref{fig:TMD1H}(a)] have been proposed as promising platforms for the occurrence of sizable BCD~\cite{you18}. 
Because of the threefold rotation symmetry of the crystalline structure, $1H$-TMDs do not satisfy the symmetry constraints discussed above. Consequently an external perturbation, namely uniaxial strain, needs to be applied. 

Generally speaking, 1H monolayers of group-VI dichalcogenides MX$_2$ (with M=Mo,W and X=S, Se, Te)  have a direct band-gap in the visible frequency range~\cite{spl10,mak10} with conduction and valence bands edges that are located at the corners of the two-dimensional hexagonal BZ. As a result, the low-energy properties of these systems are very similar to graphene but with two important differences. First, inversion symmetry is explicitly broken thus implying a non-vanishing Berry curvature~\cite{xia12,xu14}. Second, TMDs have a strong spin-orbit coupling resulting from the $d$-orbitals of the heavy metal atom. The appearance of massive Dirac cones can be immediately understood by
first neglecting spin-orbit coupling all together. Density functional theory calculations~\cite{zhu11,kad12,leb09,ata12} show that the electronic band structure consists of partially filled $d$-bands of the heavy metal ion lying between M-X $s-p$ bonding and antibonding bands. The trigonal prismatic coordination of the transition metal ions [see Fig.~\ref{fig:TMD1H}(a)] splits its $d$-bands into a single one-dimensional representation -- the $d_{z^2}$ orbital -- and two different two-dimensional representations that differ from each other in the $\pm 1$ eigenvalue of the reflection symmetry with respect to the (horizontal) mirror plane.  Specifically, the $d_{x^2 - y^2}$ and $d_{x y}$ orbitals form  the so-called irreducible $E$ representation, whereas the $d_{xz}$ and the $d_{yz}$ orbitals give rise to the so-called $E^{\prime}$ two-dimensional irreducible representation~\cite{liu13tmd}. At the $\Gamma$ point of the BZ the valence and conduction states close to the Fermi level are given by the single $d_{z^2}$ orbital and twofold degenerate states composed of the $d_{xy}$ and $d_{x^2 - y^2}$ orbitals. These two degenerate states split along the $\Gamma - K$ line thus allowing to describe the low-energy properties in terms of an effective two-band model. The states forming the effective low-energy doublet at the $K$ point are given by the combined orbitals $d_{z^2}$ and $d_{x y} \pm i d_{x^2-y^2}$ as can be simply rationalized by noticing that the little group at the $K$ point is ${\mathcal C}_{3 h}$. We emphasize that it is the absence of a vertical mirror symmetry $\sigma_{v}$ in the little group that guarantees the splitting of the $E$ doublet at the $K$ point. On the other hand, time-reversal symmetry guarantees that the states at the $K$ and $K^{\prime}$ points, which are related to each other by time-reversal symmetry, are given by $d_{x y} + i  \tau d_{x^2-y^2}$ with $\tau$ the valley index.
The effective Hamiltonian away from the $K, K^{\prime}$ of the BZ can now be derived using two-dimensional ${\bf k} \cdot {\bf p}$ theory. To do so, we notice that the threefold rotation symmetry operator can be represented as ${\mathcal C}_3 = \exp{- i \pi \sigma_z \tau_z / 3}$ whereas the time-reversal operator $\Theta= \tau_x \, {\mathcal K}$. Finally, the mirror symmetry interchanging the valleys can be simply represented as ${\mathcal M}= \tau_x$. The presence of these symmetries allows the presence of a Dirac mass $\Delta~\sigma_z/2$ where $\Delta$ corresponds to the TMD direct gap, while the triad of Pauli matrices $\sigma_{x,y,z}$ now acts in the orbital space. Using the transformation of momenta under the threefold rotation symmetry one finds to linear order in momentum a isotropic Dirac theory of the form
\begin{equation} 
{\mathcal H}^0_{TMD} = v_F \left(\tau k_x \sigma_x + k_y \sigma_y \right) + \dfrac{\Delta}{2} \sigma_z. 
\end{equation} 
The strong spin-orbit coupling of the metal $d$ orbitals require the inclusion of the atomic spin-orbit coupling ${\bf L} \cdot {\bf S}$. The ensuing effective Hamiltonian~\cite{xia12} can be then recast in the form 
\begin{equation} 
{\mathcal H}_{TMD} = {\mathcal H}^0_{TMD} - \lambda \tau \, \dfrac{\sigma_z  - 1}{2} \, s_z 
\label{eq:ham1htmd}
\end{equation}
with $s_z$ indicating the spin degrees of freedom, and $\lambda$ the halved spin splitting of the valence band edge caused by spin-orbit coupling. Note that the eigenstates of the effective ${\bf k} \cdot {\bf p}$ Hamiltonian are common eigenstates of the spin in the direction perpendicular to the 
two-dimensional TMD sheet due to the presence of the horizontal mirror symmetry ${\mathcal M}_z$. 
The low-energy Hamiltonian in Eq.~\ref{eq:ham1htmd} corresponds to two copies (but with different gaps) of the effective Hamiltonian for the surface states of topological crystalline insulators without the tilt term. The absence of the latter is enforced by the threefold rotation symmetry, which, in turn, implies the existence of three vertical mirror symmetries, and forces the BCD to vanish. 
However, the application of uniaxial strain can lower the point group symmetry making the tilt term symmetry allowed and the BCD finite. 
The presence of a strain-induced tilt term cannot be captured in an effective ${\bf k \cdot p}$ theory considering strain terms at lowest order. 
However, a microscopic model based on {\it ab initio} derived Wannier functions with strain effects incorporated by varying the interatomic bond length~\cite{fan18} shows the appearance of a finite BCD in complete agreement with the symmetry analysis. The size of the BCD is, however, much smaller than the one theoretically predicted to occur on the surface states of SnTe. The latter has been estimated to be in the nm range~\cite{Sodemann2015}, whereas BCDs of the order of $10^{-2} \AA$ are expected in the 1H structure of both WSe$_2$~\cite{you18} [see Fig.~\ref{fig:TMD1H}(b)] and MoS$_2$~\cite{son19}.

\begin{figure}
\begin{center}
  \includegraphics[width=.6\linewidth]{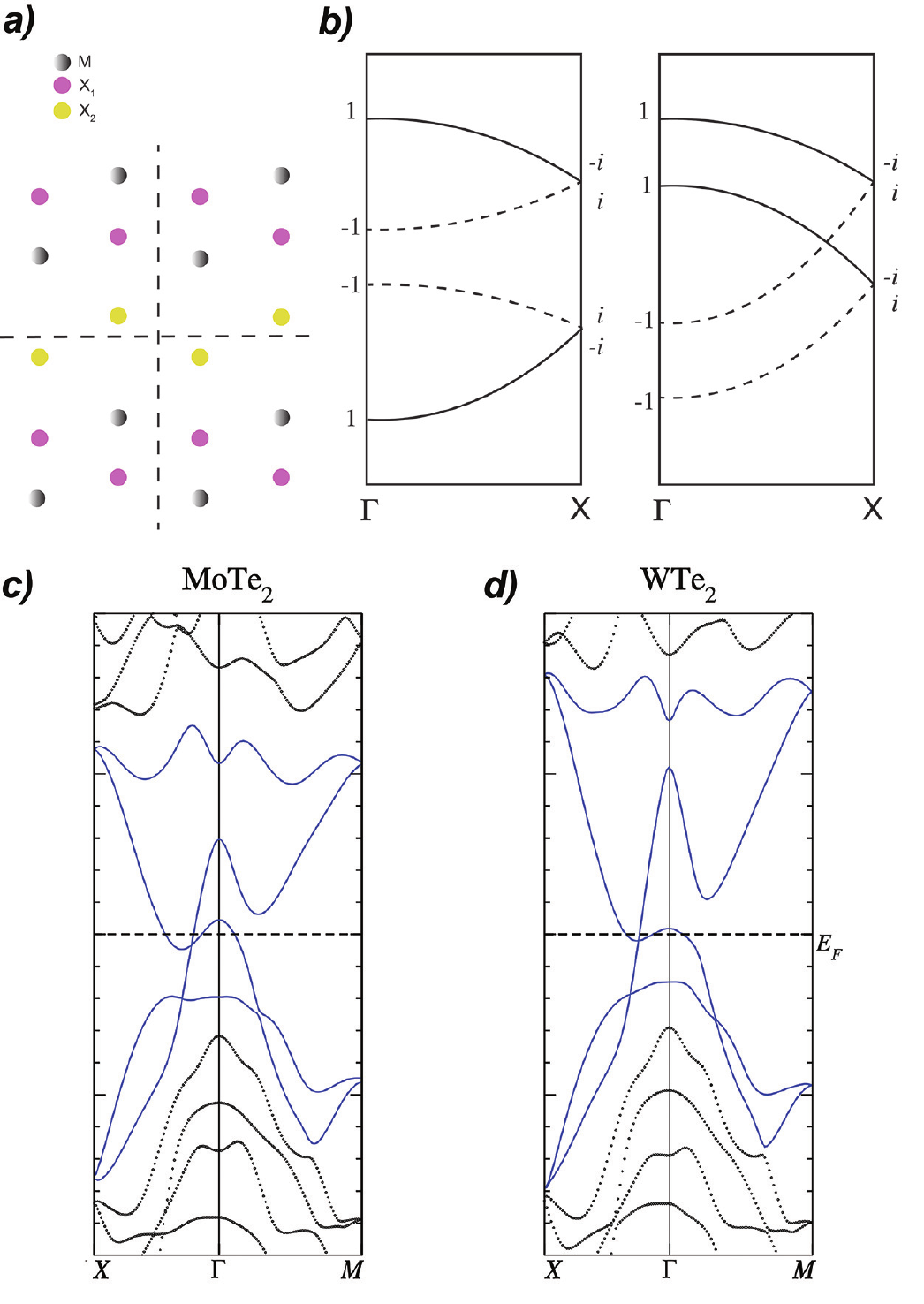}
\end{center}
  \caption{{\bf (a)} . Crystal structure of monolayer MX$_2$ in the 1T$^{\prime}$ form. The horizontal line is invariant under the screw rotation whereas the vertical line is invariant under the modified mirror symmetry. The crossing between the two dashed lines also corresponds to the center of inversion. {\bf (b)} Sketch of the possible electronic dispersions along the scew rotation symmetry $\Gamma-X$ line of the BZ. The left panel shows the band structure of a conventional insulator at even filling, whereas the right panel is a symmetry-enforced non-symmorphic semimetal. The screw rotation eigenvalues at the high-symmetry point can be used to define a topological indicator. {\bf (c),(d)} Density functional theory band structures of MoTe$_2$ and WTe$_2$ calculated without spin-orbit coupling. Panels {\bf (c),(d)} are reproduced with permission.\textsuperscript{[Ref.~\cite{mue16}]} Copyright 2016, American Physical Society.}
  \label{fig:TMD1Td}
\end{figure}

Although the hexagonal structure is the most energetically favorable one, TMD monolayers are present also in a different structural form: the so-called 1T$_d$ phase. 
It corresponds to the bulk $T_d$ phase in which WTe$_2$ and MoTe$_2$ realize Weyl semimetals~\cite{sol15,hua16,den16,jia17}. The monolayer 1T$_d$ phase intrinsically breaks bulk inversion symmetry. However, 
at sufficiently low-temperature the monolayers undergo a slight structural distortion thanks to which inversion symmetry is recovered. This structure is the so-called 1T$^{\prime}$ structure, which, as we now review, possess gapless tilted Dirac cones 
if spin-orbit coupling is removed all together. 
To show this, let us consider the relevant crystalline symmetries~\cite{mue16} of the 1T$^{\prime}$ structure [see Fig.~\ref{fig:TMD1Td}(a)]: a modified mirror symmetry $\bar{\mathcal M}_x$ corresponding to a conventional mirror line symmetry ${\mathcal M}_x$ followed by a translation by half of a lattice vector $t\left({\bf e}_x/2\right)$, and a screw rotation $\bar{\mathcal C}_{2 x}$ that is the product of a twofold rotation ${\mathcal C}_{2 x}$ around the $\hat{x}$ axis followed by the same fractional translation $t\left({\bf e}_x/2\right)$. Note that the combined presence of these two symmetries implies the presence of the inversion symmetry ${\mathcal I}=\bar{\mathcal C}_{2 x} \times \bar{\mathcal M}_x$. In the two-dimensional BZ of the system, there are two high-symmetry lines, {\it i.e.} $k_y=0$ and $k_y=\pi$ where the bands can be labelled by the eigenvalues of $\bar{\mathcal C}_{2 x}$. The latter fall into two momentum-dependent branches $\pm e^{-i k_x / 2}$ as follows from the fact that $\left(\bar{\mathcal C}_{2 x}\right)^2  = t\left({\bf e}_x\right)=e^{-i k_x}$. At the inversion-symmetric momenta $X=\left\{\pi , 0\right\}$ and $M=\left\{\pi , \pi\right\}$ the states with complex-conjugate screw rotation eigenvalues $\pm i$ are paired up by the effective ``spinless" time-reversal symmetry $\Theta={\mathcal K}$. At the other high-symmetry momenta, {\it i.e.} $\Gamma=\left\{0, 0 \right\}$ and $Y=\left\{0, \pi \right\}$ the screw rotation eigenvalues are instead real $\pm 1$. As a result, time-reversal symmetry does not imply any additional degeneracy. For even fillings, these symmetry properties can lead to a symmetry-enforced ``topological" semimetal along the $\Gamma-X$ (or $Y-M$) screw line [see the sketch in Fig.~\ref{fig:TMD1Td}(b)]. Close to the Dirac point on the screw line the linear dispersion 
is completely anisotropic. In addition, a tilt term $k_x \sigma_0$ is symmetry allowed. Density functional theory band structures in the absence of spin-orbit coupling [see Fig.~\ref{fig:TMD1Td}(c),(d)] confirmed the presence of tilted massless Dirac cones both in MoTe$_2$ and WTe$_2$~\cite{mue16}. 
This topological semimetal protected by screw rotation becomes a quantum spin-Hall insulating state when the intrinsic spin-orbit coupling is taken into account~\cite{qia14,fei17,tan17,jia17b,wu18}. In this case, the Berry curvature is identically zero due to the concomitant presence of inversion and time-reversal symmetry. However, breaking inversion symmetry via either an externally applied electric field or considering a structural distortion to the non-centrosymmetric 1T$_d$ phase, all the conditions for the appearance of a BCD are met. 
Both these situations have been theoretically studied using density functional theory calculations~\cite{you18,zha18}. 
In particular, it has been shown that WTe$_2$ in the 1T$_d$ possess BCDs of the order of $10^{-1} \AA$. Larger values have been instead found using the tunability provided by the external electric field in the 1T$^{\prime}$ phase both in WTe$_2$ and MoTe$_2$ monolayers. 
Even more importantly, a BCD of the order of $10^0 \AA$ has been estimated using circular photogalvanic measurements in WTe$_2$~\cite{xu18}. Similarly to the density functional theory calculation studies, these experiments have highlighted a large tunability of the BCD. 

\subsection{Bilayer transition metal dichalcogenides}

\begin{figure}
\begin{center}
  \includegraphics[width=.7\linewidth]{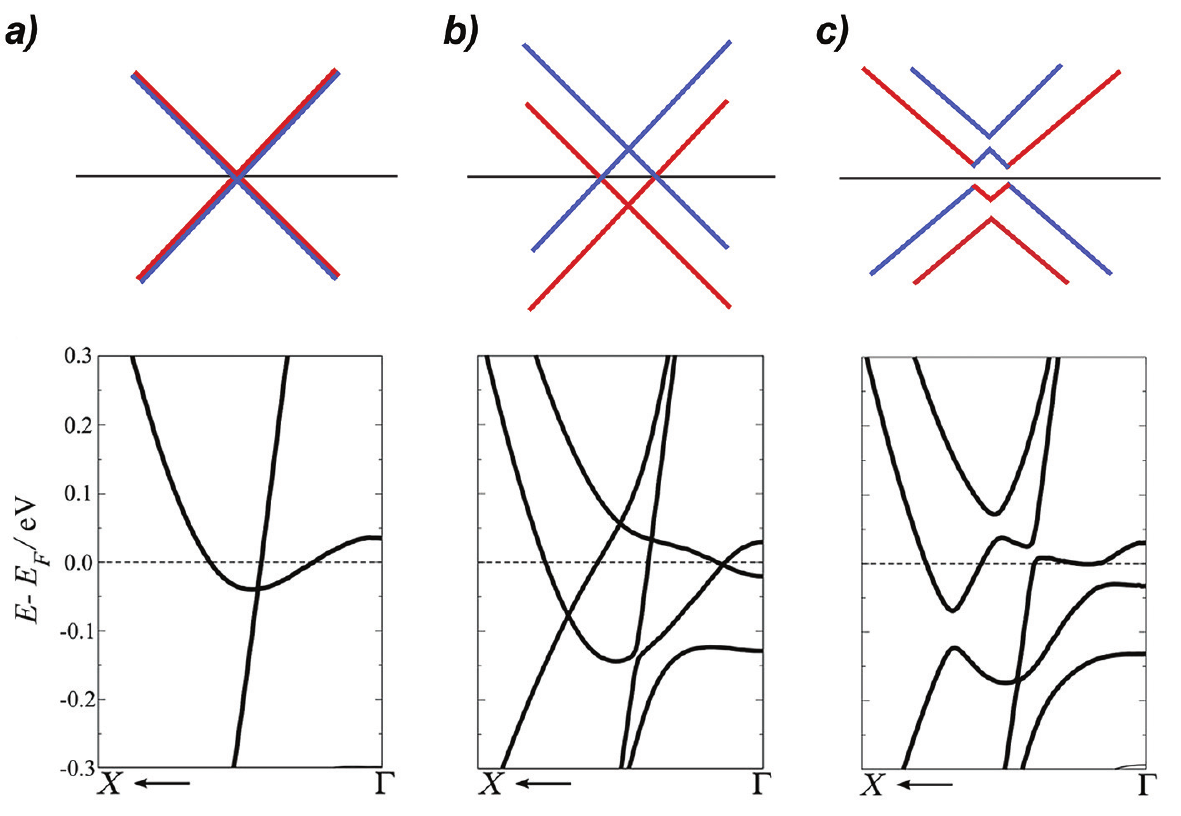}
\end{center}
  \caption{Sketch (top panels) and density functional theory band structures (bottom panels) of WTe$_2$ assuming a vanishing interlayer coupling {\bf (a)}, an hypothetical bilayer structure preserving the screw rotation symmetry {\bf (b)} and the actual stacking that breaks the screw rotation symmetry and yields massive Dirac cones {\bf (c)}. The density functional theory band structures are reproduced with permission.\textsuperscript{[Ref.~\cite{mue16}]} Copyright 2016, American Physical Society. }
  \label{fig:TMDbi}
\end{figure}

Sizable BCDs have been theoretically predicted~\cite{du18} and experimentally verified in bilayer~\cite{ma19}  and few-layer~\cite{kan19} WTe$_2$. TMD bilayers, in particular, display dipoles of the order of a few nanometers, one order of magnitude larger than the values predicted and experimentally observed in the electrically activated 1T$^{\prime}$ phase of TMD monolayers, and four order magnitudes than the one predicted in 1H monolayers. Moreover the size of the dipole can be additionally tuned with the application of an out-of-plane electric field.
The origin of such a large BCD in TMD bilayers can be understood by considering a simple mechanics of hybridization between Dirac fermions appearing in two isolated monolayers~\cite{mue16}. Assume for simplicity that on the screw lines of the TMD monolayers the Dirac cones have a vanishingly small tilt term. In the unrealistic case in which the two layers are completely decoupled the bilayer will then feature two massless Dirac cones one on top of each other [see Fig.~\ref{fig:TMDbi}(a)]. Consequently, the electronic band structure of the bilayer system will be twofold degenerate at all wavevectors. 
Next, we can assume that the two layers are coupled but stacked without any relative displacement in the $x-y$ plane. The interlayer coupling will energetically split the Dirac cones, and thus remove the twofold degeneracy of the bands. However, the Dirac points on the screw lines will still be present [see Fig.~\ref{fig:TMDbi}(b)] since the screw rotation is preserved. Finally, one can consider the stacking experimentally realized in, {\it e.g.} WTe$_2$. The relative displacement between the two layers breaks the screw rotation symmetry. Therefore, the Dirac cones acquire a mass and Berry curvature [see Fig.~\ref{fig:TMDbi}(c)]. Moreover, since the original Dirac cones have opposite chirality the Berry curvature of the resulting massive Dirac cones are opposite. After they become gapped, these untilted Dirac cones yield a strong BCD since the right moving states and the left moving states are characterized by an opposite Berry curvature. Adding now the effect of the spin-orbit coupling and the original tilt of the Dirac cones does not change qualitatively any of the conclusion. 

\section{Berry curvature dipole in the absence of spin-orbit coupling} 
\label{sec:nosoc2d}
The mechanism at work in bilayer TMDs does not strictly require the presence of titled massive Dirac cones, neither of a sizable spin-orbit coupling. This feature highlights the possibility to observe the non-linear Hall effect even in two-dimensional materials made of light elements. Graphene has been consequently put forward as a paradigmatic example of spin-orbit-free materials  a non-vanishing BCD~\cite{bat19}. Importantly, signatures of a non-linear Hall effect with time-reversal symmetry have been recently found in strained bilayer graphene~\cite{ho21}. As we will discuss below, the origin of the non-linear Hall effect in graphene does not stem from the Dirac cone shifting mechanism of bilayer WTe$_2$. It is instead the warping of the Fermi surface the physical characteristic, which is responsible for the non-vanishing of the BCD. 

\subsection{Uniaxially strained monolayer graphene}

The wallpaper group of monolayer graphene is $p6mm$. It is generated by the point group ${\mathcal C}_{6v}$ and in-plane translations. The point group is generated by a threefold rotation symmetry, a mirror symmetry and a twofold rotation. Since spin-orbit coupling can be neglected all together in graphene, the twofold rotation coincides with inversion symmetry. As a result, the Berry curvature identically vanishes. 
A possible way to open up a gap and break inversion symmetry is to induce a sublattice imbalance~\cite{sem84}, {\it i.e.} a charge density wave instability. This can be achieved for instance by placing graphene on a lattice-matched substrate. Density functional theory calculations~\cite{gio07} first identified hexagonal boron nitride as a possible substrate inducing a band gap of  the order of $25$~meV. 
The small nominal lattice mismatch between graphene and hexagonal boron nitride can also lead to a large moire' superlattice characterized by the generation of mini Dirac cones at the expense of a full  band gap opening~\cite{ort12,yan12,pon13,wal13}. 
Signatures of a commensurate-incommensurate transition for graphene on top of hexagonal boron nitride~\cite{woo14} have revealed that both these situations can occur in practice: moire' gapless regions are separated by the gapped commensurate ones of interest for our discussion below. 
As for the case of TMD monolayers in the 1H phase, the point group ${\mathcal C}_{3v}$ of gapped graphene does not allow for a non-vanishing BCD. However, the application of uniaxial strain lowers the symmetry of the system to  
${\mathcal C}_v$; the presence of a single mirror line then allows for the onset of a non-linear Hall effect. Let us now examine the electronic characteristic of a graphene layer in the presence of uniaxial strain. 
To do so, we start out from the graphene tight-binding Hamiltonian~\cite{cas09} in its simplest form, {\it i.e.} considering only hopping processes between nearest neighbor atomic sites. It reads: 
\begin{equation}
{\mathcal H}_{MLG}= -  \sum_{i \, n} t_n a^{\dagger}_i b_{i + \delta_n} + {\it c.c.} - \dfrac{\Delta}{2} \sum_{i}  a^{\dagger}_i a_i + \dfrac{\Delta}{2} \sum_{i} b^{\dagger}_i b_i, 
\end{equation} 
where $a^{\dagger}$ and $b^{\dagger}$ ($a$, $b$) are creation (annihilation) operators on the A and B sublattices respectively. In the equation above, $\Delta$ indicates the substrate-induced band gap, the subscript $i$ runs over all unit cell positions, and we introduced the three nearest neighbor vectors
\begin{equation}
{\boldsymbol \delta}_{1}=\dfrac{a}{\sqrt{3}} \left\{\dfrac{\sqrt{3}}{2}, \dfrac{1}{2} \right\} \hspace{.5cm} {\boldsymbol \delta}_2= \dfrac{a}{\sqrt{3}}  \left\{\dfrac{-\sqrt{3}}{2}, \dfrac{1}{2} \right\} \hspace{.5cm} {\boldsymbol \delta}_3 = \dfrac{a}{\sqrt{3}} \left\{0, -1 \right\}. 
\end{equation} 

\begin{figure}
\begin{center}
  \includegraphics[width=.7\linewidth]{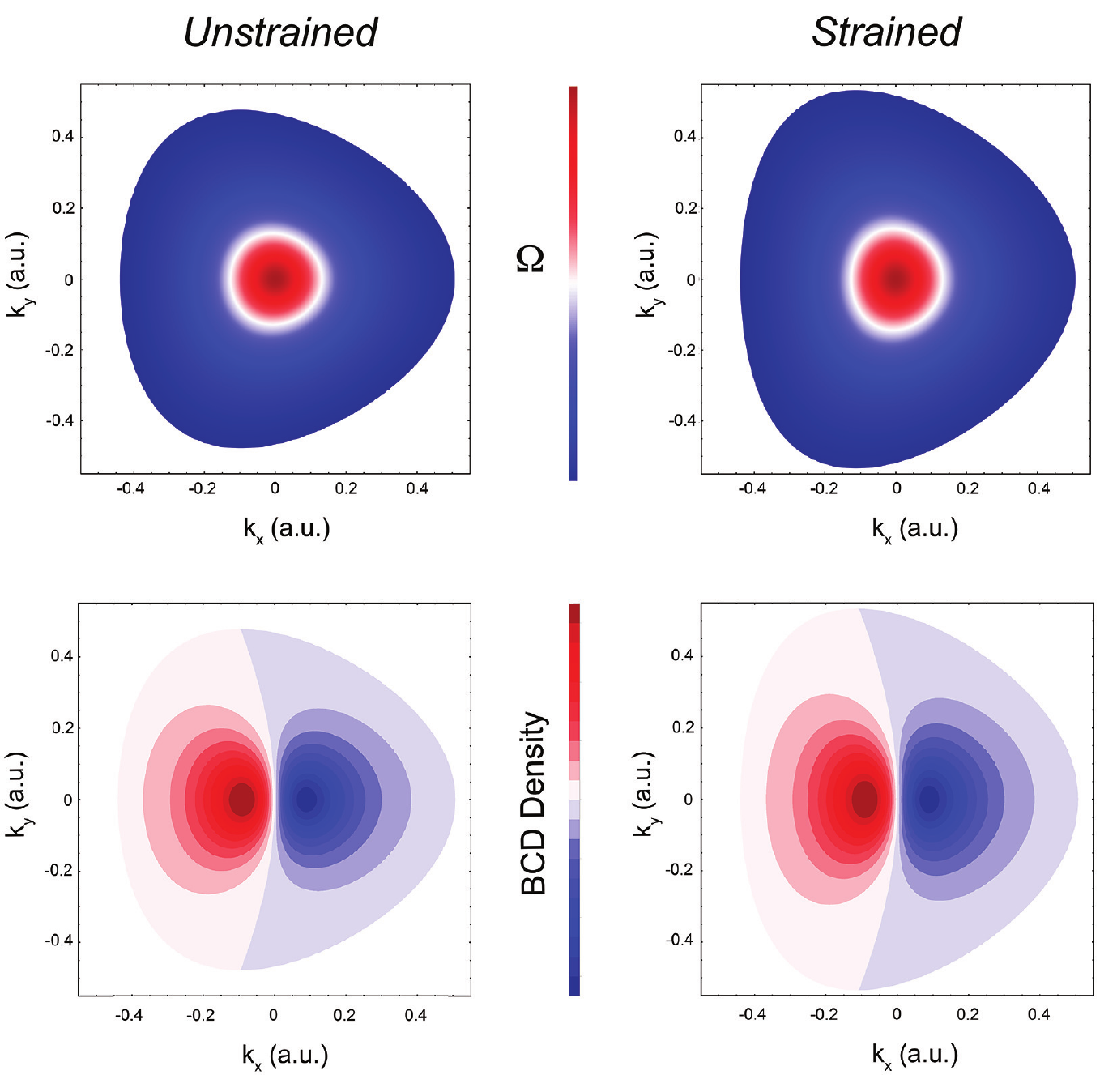}
\end{center}
\caption{Berry curvature and dipole density for unstrained (left panels) and strained (right panels) monolayer graphene. When strain is present the deformation of the Fermi surface leads to a finite value of the dipole. Note that we have considered for simplicity the trigonal warping term that respects the threefold rotation symmetry, {\it i.e.} $\lambda_{1}=\lambda_{2}=\lambda_{3}$. The distortion of the Fermi surface is therefore entirely due to the anisotropic Fermi velocity.}
\label{fig:MLG}
\end{figure}

Furthermore, the presence of strain implies that the hopping amplitudes $t_n$ explicitly depend on the nearest neighbor vectors. Specifically,
$t_n = t_0 \left( 1 - \beta \, \delta u_n \right)$
where $t_0$ is the hopping amplitude for the pristine threefold rotation symmetric honeycomb lattice, the lattice parameter $\beta$ can be determined by Raman spectroscopy whereas the relative distance changes $\delta u_n$ can be expressed in terms of the strain tensor components $\epsilon_{i\, j}$ as 
\begin{equation}
\delta u_n = \dfrac{\delta_n^i \, \delta_n^j}{a^2} \epsilon_{i j}.
\end{equation}
 To proceed further, we go to momentum space and write the Bloch Hamiltonian as
\begin{equation} 
{\mathcal H}_{MLG} = - \sum_n t_n \left( \begin{array}{cc} \dfrac{\Delta}{2} & e^{-i \left({\bf K}^{(\prime)} + {\bf q} \right) \cdot {\boldsymbol \delta}_n} \\  e^{i \left({\bf K}^{(\prime)} + {\bf q} \right) \cdot {\boldsymbol \delta}_n}  & - \dfrac{\Delta}{2} 
 \end{array} \right), 
\end{equation}
where we have rewritten the momenta as ${\bf k} = {\bf K}^{(\prime)} + {\bf q}$ since we are interested in the electronic properties close to the ${\bf K}$ or ${\bf K}^{\prime}$ valleys of the BZ given by ${\bf K}= \left\{ \dfrac{4 \pi}{3 \sqrt{3} a} , 0 \right\}$ and ${\bf K}^{\prime}= \left\{-\dfrac{4 \pi}{3 \sqrt{3} a} , 0 \right\}$. The Bloch Hamiltonian can be then expanded to linear order in the small momenta $q$. Using simple vector identities and assuming an anisotropic biaxial strain with $\epsilon_{xx} \neq \epsilon_{yy} \neq 0$ and $\epsilon_{x y} \equiv 0$ the continuum low-energy Hamiltonian near the two valleys of the BZ can be then recast as 
\begin{equation}
\mathcal{H}_{eff}(\mathbf{q})= \xi v_x q_x\sigma_x+v_F {\mathcal A}_x\sigma_x + v_y q_y \sigma_y+ \frac{\Delta}{2}\sigma_z
\end{equation}
where $\xi=\pm 1$ is the valley index. In the equation above, ${\mathcal A}_x = \sqrt{3}\, \beta(\epsilon_{xx}-\epsilon_{yy})/ (2 a)$ is the well-known strain-induced ``pseudo"-gauge field~\cite{gui10,voz10,gop12} whereas $v_F=\sqrt{3}t_0 a / 2$  is the Fermi velocity of the Dirac carriers in unstrained samples. In addition, $v_{x}= v_F[1-\beta  (3\epsilon_{xx}+\epsilon_{yy})/4]$ and $v_y=v_F[1-\beta (\epsilon_{xx}+3\epsilon_{yy})/4]$ are renormalized Fermi velocities that become anisotropic when explicitly considering the momentum-strain coupling~\cite{dej12}. The Dirac cones explicitly acquire an anisotropic character, consistent with the reduction of the point group symmetry. The warping of the Fermi surface can be instead captured by keeping terms up to quadratic order in the momentum $q^2$. They read $\Delta \mathcal{H}_{eff}(\mathbf{q})= (\lambda_1 q_y^2-\lambda_2 q_x^2)\sigma_x+2\xi q_xq_y\lambda_3\sigma_y$, 
where the warping coefficients $\lambda_{1,2,3}$ are explicitly renormalized by strain. In the absence of it, $\lambda_{1} \equiv \lambda_{2} \equiv \lambda_{3}$ and we reach the trigonal warping of pristine graphene~\cite{cas09}. 
By limiting to homogeneous strain, the presence of the pseudo-gauge field can be simply reabsorbed by a proper redefinition of the small momentum: The pseudo-gauge field in fact only shifts the Dirac cones away from the ${\bf K}^{(\prime)}$ valley. The anisotropy in the Fermi velocity combined with the warping of the Fermi surface instead produces a distortion of the Fermi lines [see Fig.~\ref{fig:MLG} ] that endows the system with a non-zero BCD even if an explicit tilt term is absent~\cite{bat19}. 
Precisely as for the case of TMDs both the gapped Dirac cones contribute in an equal manner to the total BCD~\cite{Sodemann2015}. 
Note also that a finite value of the BCD occurs even if the strain-induced renormalization of the warping coefficients is disregarded [c.f. Fig.~\ref{fig:MLG}] . This finite BCD is pinned to the direction orthogonal to the surviving mirror line of the system, and thus parallel to the zigzag direction of the honeycomb lattice. For electronic densities $n_{el} \sim 10^{10}~$cm$^{-2}$, and a substrate-induced gap $\Delta \simeq 20~$meV, the size of the BCD has been calculated to lie in the $10^{-3}$~nm range assuming at 5~\%  strain. This is the same order of magnitude of the BCD in the 1H phase of monolayer TMDs discussed above.

\subsection{Strained bilayer graphene}

\begin{figure}
\begin{center}
  \includegraphics[width=.6\linewidth]{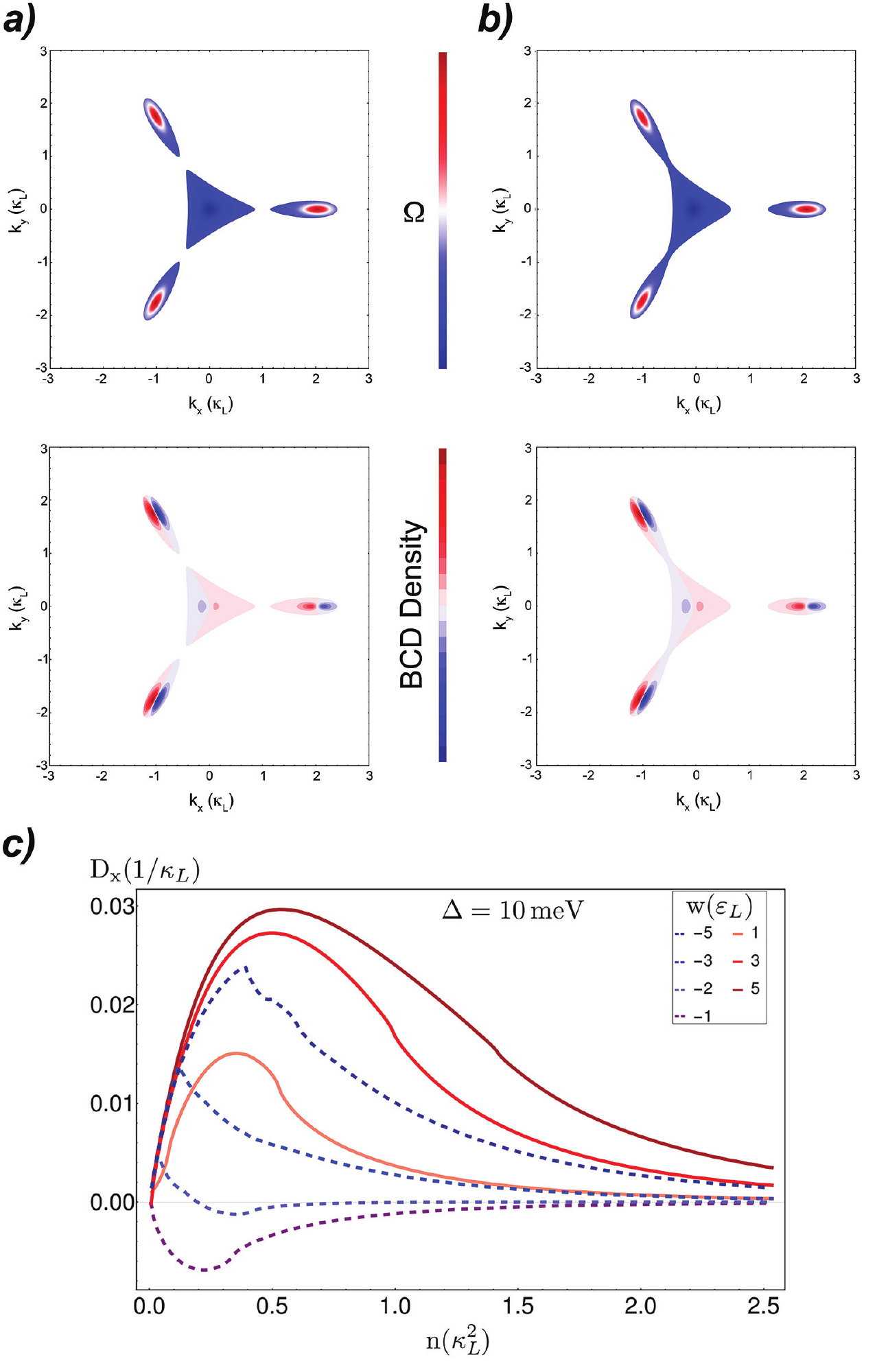}
\end{center}
  \caption{{\bf (a)} Berry curvature and Berry curvature dipole density in gated bilayer graphene in the absence of strain below the Lifshitz transition. {\bf (b)} Same with homogeneous uniaxial strain. Beside yielding a non-vanishing Berry curvature dipole, the uniaxial strain also promotes a Lifshitz transition. {\bf (c)}  Berry curvature dipole in bilayer graphene assuming a gate-induced gap $\Delta = 10$~meV and various strains as a function of the carrier density measured in units of $\kappa_L^2$ with $\kappa_L =0.035$nm$^{-1}$. Panel {\bf (c)} is reproduced with permission.\textsuperscript{[Ref.~\cite{bat19}]} Copyright 2019, American Physical Society. }
  \label{fig:BLG}
\end{figure}

The appearance of a non-vanishing BCD in the complete absence of spin-orbit coupling and tilted Dirac cones is not specific of monolayer graphene. As we review below, a finite non-linear Hall voltage can appear in gated bilayer graphene subject to mechanical deformations. Even more importantly, the BCD has been shown to lie in the nm range and therefore comparable with the one of bilayer WTe$_2$. To understand how such a large BCD appears in gated bilayer graphene, we first discuss strain effects on the electronic properties in Bernal stacked bilayer graphene. To do so, we start out with the effective continuum low-energy Hamiltonian~\cite{McCann2006,Mucha-Kruczynski2010,McCann2013} for the electrons on the A$_{1,2}$ and B$_{1,2}$ sublattices of the two $1,2$ carbon sheets. It reads: 
\begin{equation}
{\mathcal H}_{\mathrm{4\times 4}}=
\begin{pmatrix}
-\frac{\Delta}{2} &  v_0 \pi^\dagger + \mathcal{A}_0 & 0 & v_3 \pi + \mathcal{A}_3\\
v_0 \pi + \mathcal{A}_0 & -\frac{\Delta}{2} & \gamma_1 & 0 \\
0 & \gamma_1 & \frac{\Delta}{2} &  v_0 \pi^\dagger + \mathcal{A}_0 \\
v_3 \pi^\dagger + \mathcal{A}_3& 0 &  v_0 \pi + \mathcal{A}_0 & \frac{\Delta}{2}
\end{pmatrix}
\label{eq:Hss}
\end{equation}
where $\pi = k_x + i k_y$ and $v_{0/3}$ are two strain independent parameters with the dimension of a velocity. Specifically, $v_0$ is the Fermi velocity of the Dirac electrons in each monolayer and is thus related to the intralayer hopping amplitude $\gamma_0$. By contrast, the velocity $v_3$ is related to the so-called ``skew" interlayer hopping amplitude $\gamma_3$ connecting the $A_{1}$ and $B_{2}$ sites. The $A_{2}$ and $B_{1}$ lie on top of each other and are coupled by the intercoupling $\gamma_{1}$. The effects of strain appear as two pseudo-gauge fields $\mathcal{A}_{0/3}$~\cite{Mucha-Kruczynski2011}. Assuming the stresses are applied along the principal crystallographic directions, the pseudogauge fields $\mathcal{A}_{0/3}=\frac{3}{4}\gamma_{0/3} (\epsilon_{xx}-\epsilon_{yy})\beta_{0/3}$ with $\beta_{0/3}$ two different dimensionless parameters related to the elastic properties of the material. Finally, the different on-site potential in the two layers $\pm \Delta / 2$ is due to an externally applied electric field~\cite{Oostinga2008}. For energies near the Fermi level the states coupled by $\gamma_1$ can be eliminated
through a Schrieffer-Wolff transformation~\cite{Schrieffer1966}. After a gauge transformation $k_x \rightarrow k_x -\mathcal{A}_0$, 
one obtains the effective $2\times2$ Hamiltonian, 
\begin{align}
{\mathcal H}_{b}= \left[ -\frac{1}{2m} (k_x^2 - k_y^2) + \xi v_3 k_x  + w\right]\sigma_x  \\ \nonumber \label{eq:Hdip}
-\left(\frac{1}{m}k_xk_y + \xi v_3 \right)\sigma_y + \frac{\Delta}{2}\sigma_z,
\end{align}
where we introduced the effective mass $m= \gamma_1/2v_0^2$ and the strain coupling $w=\frac{3}{4}\gamma_3 (\epsilon_{xx}-\epsilon_{yy})(\beta_3-\beta_0)$. The inclusion of the skew interlayer hopping has a twofold effect. First, it gives an explicit trigonal warping to the Fermi surface. Second, and most important, it changes the topology of the Fermi surface at low energies. In the absence of strain for instance, the low-energy quadratic band crossing point occurring for $v_3=0$ is changed in favour of the appearance of four Dirac cones, one of which is located at the ${\bf K}$ and ${\bf K}^{\prime}$ points of the BZ. The remaining three ``leg" Dirac cones, with a Berry phase opposite to the central cone one, are arranged in a three-fold rotation symmetric fashion [see Fig.~\ref{fig:BLG}(a)] and have a distance in momentum given by $\kappa_L = m v_3$. The Lifshitz transition where the Fermi surface changes topology~\cite{McCann2013} occurs at $\epsilon_L=m v_3^2 / 2$ in the $\Delta = 0$ inversion-symmetric case. 

The appearance of a non-vanishing BCD in the system can be understood considering the Berry curvature properties already in this strain-free ($w=0$) case~\cite{bat19}. As shown in Fig.~\ref{fig:BLG}(a) the central Dirac cone has a ${\mathcal C}_3$ symmetric Berry curvature and thus it gives a vanishing contribution to the BCD. When taken by themselves, the leg Dirac cones instead have only a mirror symmetric Berry curvature profile and therefore their contribution to the BCD is non-zero. This is because each of the leg gapped Dirac cones is described by the low-energy Hamiltonian introduced for monolayer graphene with a warping term such that $\lambda_{1}=\lambda_{2}=\lambda_{3}$. However, the contributions of the leg Dirac cones cancel each other due to the threefold rotation symmetry. Once strain is introduced, the perfect cancellation of the leg Dirac cone contributions is lost and a finite overall BCD appears. In addition, the central massive Dirac cone also yields a non-zero contribution to the BCD for $w \neq 0$.  As explicitly shown in Fig.~\ref{fig:BLG}(b) strain promotes changes in the topology of the Fermi surface and yield additional Lifshitz transition. Moreover, it can also lead to Dirac cone annihilation processes as shown for the inversion symmetric case in Ref. 
Independent of the specific fermiology, the system always develops a sizable total BCD. The existence of various Lifshitz transition is imprinted in the presence of cusp and inflection points in the behavior of the BCD as a function of the carrier density [see Fig.~\ref{fig:BLG}(c)]. Importantly for a strain term $w= - 5 \epsilon_L$, which corresponds roughly to a 1\% strain, and a for gate-induced gap $\Delta = 10~$meV, the BCD $D_x \simeq 1$~nm assuming a carrier density $n \simeq 10^{11}~$cm$^{-2}$~\cite{Oostinga2008}. This value, which, as mentioned above, is comparable to the one experimentally found in bilayer WTe$_2$ can be boosted by an order of magnitude for smaller values of the gate-induced gap. And indeed dipoles even in the tens of nanometer scale have been recently found in the strained bilayer graphene nanoarchitectures of Ref.~\cite{ho21}. 

\section{Three-dimensional materials}
\label{sec:3dmaterials}
\begin{figure}
\begin{center}
  \includegraphics[width=.73\linewidth]{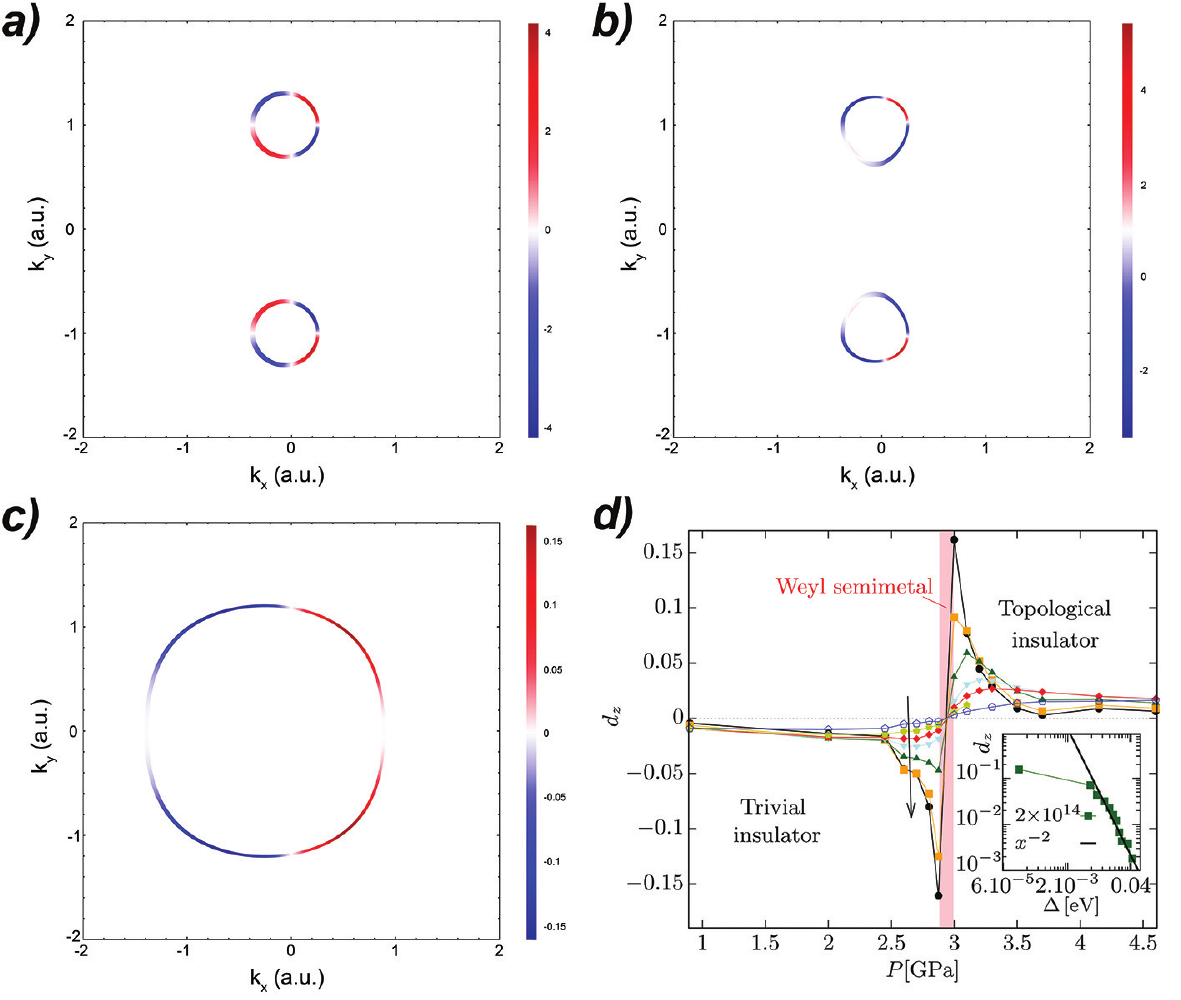}
\end{center}
  \caption{{\bf (a)} Berry curvature dipole line density $\Omega_{{\bf k} x} {\bf v}_{{\bf k}}^y$ evaluated on the $k_z=0$ plane for a pair of Weyl nodes using the effective ${\bf k \cdot p}$ Hamiltonian for isolated cones. The dipole $D_{x y}$ clearly vanishes by symmetry. {\bf (b)} Same using the model Eq.~\ref{eq:weyl2} for the creation of a pair of Weyl nodes. In this case there isn't any symmetry forcing the dipole to be zero. {\bf (c)} Berry curvature dipole line density on the $k_z=0$ plane for the model Eq.~\ref{eq:weyl2} in the insulating regime. Provided there is a finite Fermi surface, the Berry curvature dipole is non-vanishing. {\bf (d)} Berry curvature dipole as a function of pressure in BiTeI as obtained using density functional theory calculations. The inset shows the behavior in the topological insulating region while approaching the topological phase transition to a Weyl semimetal. Panel {\bf (d)} is reproduced with permission.\textsuperscript{[Ref.~\cite{fac18}]} Copyright 2018, American Physical Society. }
  \label{fig:weyl}
\end{figure}

Although the non-linear Hall effect in time-reversal symmetric conditions finds its natural realization in two-dimensional materials, it can also arise in three-dimensional bulk crystals. TaIrTe$_4$, a type-II Weyl semimetal ternary compound~\cite{koe16,hau17,bel17}, has been recently shown to possess a non-linear Hall effect surviving even at room temperature~\cite{kum21}. Generally speaking, the fact that a finite BCD can appear in Weyl semimetals with tilted cones~\cite{mat19,zen20,sin20}, be them either of type-I or of strongly overtilted type-II, can be understood considering ${\bf k} \cdot {\bf p}$ theory. Let us analyze, for simplicity, a Weyl semimetal with an isotropic Fermi velocity. Its effective low-energy Hamiltonian reads
\begin{equation}
{\mathcal H}_{Weyl}= \chi v_0 \sum_{i=x,y,z} k_i \sigma_i + u_x k_x~{\mathcal I}, 
\end{equation}
where $\chi= \pm 1$ is the chirality or the topological charge of the Weyl cone~\cite{arm18}, and the tilt direction has been chosen to be the ${\hat z}$ direction. This continuum theory describes a type-II Weyl semimetal for $\left| u_x / v_0 \right| >1 $, whereas for $\left| u_x / v_0 \right| < 1 $ the cone is of the type-I form. 
Since the total topological charge in the full BZ of a generic system has to vanish, each Weyl cone has to come with a partner of opposite chirality. In addition, time-reversal symmetry implies that each Weyl cone has a time-reversed partner of the same chirality. The total number of Weyl cones in a time-reversal symmetric topological semimetal is therefore $4 n$, with $n$ integer. It is possible to show that
each Weyl cone yields a finite BCD. This can be seen already in the untilted $u_x \equiv 0$ limit. 
Using that the Berry curvature of a single Weyl cone is ${\boldsymbol \Omega} = \chi {\bf k} / (2  \left|{\bf k}\right|^3)$ and that the carrier velocity is ${\bf v} = v_0 \hat{\bf k}$, the components of the BCD pseudotensor read as 
\begin{equation}
D_{xx} = D_{yy}=D_{zz}= \dfrac{\chi}{3}~\dfrac{1}{4 \pi^2}, 
\label{eq:BCDweyl}
\end{equation}
whereas the off-diagonal components identically vanish. We emphasize that there are inconsistencies in the literature as it concerns the BCD of a single Weyl cone~\cite{ros18}. We refer the reader to Ref.~\cite{mat19} for a discussion on this point. 
With the BCD that depends exclusively on the topological charge of the Weyl nodes, it immediately follows that in the absence of a tilt term the $D_{ii}$'s summed over all nodes of a Weyl semimetal are identically zero. 
Assuming instead a sizable tilt term, the situation changes qualitatively. The diagonal non-vanishing components of the BCD indeed acquire a specific dependence on the dimensionless parameter $\delta = u_z / v_0$. This has been evaluated in Ref. and reads 
\begin{eqnarray} 
D_{yy}=D_{zz}&=&\dfrac{1}{8 \pi^2} \dfrac{1}{\delta^3} \left( \delta + \dfrac{\delta^2 - 1}{2} \log{\dfrac{1+ \delta}{1- \delta}} \right)    \nonumber \\ 
D_{xx}&=& \dfrac{1}{4 \pi^2} \dfrac{\delta^2 - 1}{\delta^3} \left( \delta - \dfrac{1}{2} \log{\dfrac{1+ \delta}{1- \delta}} \right). 
\end{eqnarray}
The equations above reduce to Eq.~\ref{eq:BCDweyl} in the $\delta \rightarrow 0$ limit. In addition, it is interesting to note that, independent of the tilt term, the trace of the BCD pseudotensor is a universal quantity $\Tr{D} = \chi / (4 \pi^2)$.  
As in the untitled case, the off-diagonal components of the pseudotensor vanish as explicitly shown in Fig.~\ref{fig:weyl}(a). 
The $\delta$-dependence of the dipole components $D_{ii}$ can make the total BCD finite in a generic Weyl semimetal. This is because the dipole tensors of the Weyl nodes related by time-reversal symmetry are same and are 
not entirely cancelled by the tensors corresponding to the Weyl nodes with opposite topological charge. This occurs in materials, such as SrSi$_2$~\cite{hua16b}, where Weyl nodes of opposite topological charge are not mapped into each other by point group symmetries. According to this analysis, however, materials with additional mirror symmetries, TaAs being a paradigmatic example~\cite{hua15,wen15,xu15,lv15}, are forced to have a vanishing total BCD for the simple reason that Weyl nodes of opposite charge are mapped by the mirror symmetry, and therefore constrained to have opposite Berry curvatures.   

Although this feature is in agreement with the fact that a three-dimensional material with mirror symmetries cannot have a symmetric BCD pseudotensor, it does not justify why the antisymmetric part of the BCD should be identically zero.  Density functional theory calculations  have shown that TaAs has a large dipole~\cite{zha18b} with $D_{xy} = - D_{yx} \neq 0$ that is symmetry-allowed. The resolution of this paradox comes from the fact that higher-order terms in momentum ${\bf k}$ can change the properties of the Berry curvature. This can be demonstrated considering a ${\bf k} \cdot {\bf p}$ model for a pair of Weyl nodes in the presence of a mirror symmetry~\cite{fac18}.  
It reads
\begin{equation}
{\mathcal H}= v_x k_x \sigma_x + v_z k_z \sigma_z + \dfrac{k_y^2 - \lambda}{2 m} \sigma_y + u_x k_x {\mathcal I}. 
\label{eq:weyl2}
\end{equation} 
This models predicts the presence of two Weyl nodes at $k_y = \pm \sqrt{\lambda}$ with a corresponding Fermi velocity $v_y= \sqrt{\lambda / m}$. Fig.~\ref{fig:weyl}(b) shows the corresponding $D_{x y}$ dipole surface density assuming a perfectly isotropic Fermi velocity for the two Weyl cones. As compared to the result obtained using the linear Weyl effective Hamiltonian [c.f. Fig.~\ref{fig:weyl}(a)], it is clear that the model Eq.~\ref{eq:weyl2} yields a finite antisymmetric BCD [c.f. Fig.~\ref{fig:weyl}] in agreement with the density functional theory calculations results for the type-I TaAs and the type-II MoTe$_2$ Weyl semimetals. 

It is interesting to note that in the $\lambda<0$ parameter range the model in Eq.~\ref{eq:weyl2} predicts an insulating state with a gap $\Delta = \left(\left| \lambda \right| / m \right)$.  Even in this regime there is a finite dipole (surface) density antisymmetric component [see Fig.~\ref{fig:weyl}(c)], provided the system can be doped with charge carriers and thus possesses a finite Fermi surface. 
The Rashba semiconductor BiTeI~\cite{ish11} has been predicted to become a strong three-dimensional topological insulator~\cite{bah12} at moderate pressures $ >3$~GPa. These two phases are separated by an intermediate Weyl semimetal phase~\cite{liu14}. These ``topological" phase transition can be therefore captured by the model in Eq.~\ref{eq:weyl2} assuming that changing the pressure the parameter $\lambda$, or equivalently the bulk gap $\Delta$, is varied. Fig.~\ref{fig:weyl}(d) shows the behavior of the antisymmetric BCD $d_z = (D_{x y} - D_{y x})/2$ computed in a recent density functional theory calculation study~\cite{fac18} for various carrier density. It highlights how the BCD gets strongly enhanced while approaching the Weyl semimetal phase. This is in agreement with the fact that the model of Eq.~\ref{eq:weyl2} predicts a divergent antisymmetric BCD in the insulating regime when the $\Delta \rightarrow 0$ limit is taken [c.f. the inset in Fig.~\ref{fig:weyl}(d)]. This demonstrates how non-linear Hall measurements can provide a unique signature of the pressure-induced topological phase transition~\cite{qi17} in BiTeI, even if local order parameters are absent. 

\section{Conclusions}
\label{sec:conc}
In this concluding section, we summarize what has been achieved with the recent studies on the NLHE in time-reversal symmetric conditions, and also point out possible directions for future research on this recently discovered phenomenon. 
Being regulated by the Berry curvature dipole, the time-reversal symmetric NLHE can be used as a direct probe of the geometry of the Bloch states in non-magnetic systems. 
This intrinsic quantum property can be only unveiled in crystals with unusually low crystalline symmetry. 
Bilayer WTe$_2$ has been the first investigated material where such symmetry requirements are fulfilled. It has led to the direct observation and measurement~\cite{ma19} of the BCD in a pristine crystal. 
The crystalline symmetry requirements for a non-vanishing BCD can be however also achieved using the capability of externally applied fields to lower the point group symmetry of a crystal. 
Breaking inversion symmetry, externally applied electric fields, for instance, allow for a time-reversal symmetric NLHE even in crystal with a native Berry curvature that identically vanishes. 
This scheme has been successfully applied in the quantum spin Hall 1T$^{\prime}$ phase of transition metal dichalcogenides~\cite{xu18}. 
Mechanical deformations are yet another knob that can be used to decrease the crystalline symmetry of materials and eventually allow for a NLHE. Contrary to external electric fields, strain does not break the centrosymmetry of a crystal. However, it can reduce the rotational symmetry to make this effect visible.
In principle, also externally applied magnetic field could be used to meet to lower the symmetry and meet the NLHE requirements. A planar magnetic field breaks the time-reversal invariance. However, the absence of Lorentz force combined with the presence of a residual mirror symmetry would guarantee the complete absence of a linear dissipationless Hall conductance thus making the non-linear Hall effect the only present transverse resistance. This effect has been recently dubbed anomalous non-linear planar Hall effect~\cite{bat21}. 

Independent of the presence and nature of the external perturbation, the appearance of the time-reversal symmetric NLHE was originally believed to necessitate a substantial spin-orbit coupling. However, the  experimental demonstration of non-linear transverse currents  in strained (bilayer) grahene~\cite{ho21} has shown that  this effect could be also pursued in systems made of light elements,  where the presence of spin-orbit coupling could be neglected all together. This has opened new material avenues for studies of the NLHE with time-reversal symmetry. Twisted bilayer graphene -- a material structure where correlated insulating behavior~\cite{cao18} and superconductivity~\cite{cao18s} have been shown to arise -- has been recently suggested to host a substantial BCD~\cite{pan20,zha20} on the order of tens of nanometers. In twisted bilayer graphene strain with an opposite sign in the two graphene layers has been imaged in STM experiments~\cite{xie19}. Moreover, when encapsulated with hexagonal boron nitride the inversion symmetry of twisted bilayer graphene is broken. As a result, all symmetry requirements for a finite BCD are fulfilled.

We conclude by mentioning that non-linear transverse currents in the presence of time-reversal symmetry have been suggested to occur also in the strong three-dimensional topological insulators of the Bi$_2$Te$_3$ material class~\cite{zha09}. The anomalous single surface Dirac cones in these materials are characterized by an intrisic chirality that enables a novel skew scattering~\cite{he21} similar to the one predicted in unstrained bilayer graphene~\cite{iso20}. However, it is important to note that the non-linear response found in these materials does not represent a dissipationless non-linear Hall conductance, and is therefore compatible with the trigonal symmetry of the Bi$_2$Te$_3$ surface states~\cite{fu09}. Finally, non-linear responses using the topological surface states of Bi$_2$Te$_3$ have been demonstrated to occur also in the presence of external planar magnetic fields~\cite{he19}.

\begin{acknowledgements}
I acknowledge support from a VIDI grant (Project 680-47-543) financed by the Netherlands Organization for Scientific Research (NWO). 
\end{acknowledgements}

\end{document}